\documentclass{JHEP3}
\usepackage{subfigure}
\usepackage{epsfig}
\usepackage{graphicx}
\usepackage{enumerate}
\usepackage{amsmath}
\title{Holographic Renormalization of Asymptotically Flat Spacetimes}

\author{Robert B. Mann\\
Perimeter Institute for Theoretical Physics, Ontario N2J
2W9, Canada \\
and Department of Physics, University of Waterloo Waterloo,
Ontario N2L 3G1, Canada \texttt{mann@avatar.uwaterloo.ca}}

\author{Donald Marolf  \\
Physics Department, UCSB, Santa Barbara, CA 93106
\texttt{marolf@physics.ucsb.edu}}

\abstract{  A new local, covariant ``counter-term'' is used to
construct a variational principle for asymptotically flat
spacetimes in any spacetime dimension $ d \ge 4$.  The new
counter-term makes direct contact with more familiar background
subtraction procedures, but is a local algebraic function of the
boundary metric and Ricci curvature. The corresponding action
satisfies two important properties required for a proper treatment
of semi-classical issues and, in particular, to connect with any
dual non-gravitational description of asymptotically flat space.
These properties are that  1) the action is finite on-shell and 2)
asymptotically flat solutions are stationary points under {\it
all} variations preserving asymptotic flatness; i.e., not just
under variations of compact support.  Our definition of asymptotic
flatness is sufficiently general to allow the magentic part of the
Weyl tensor to be of the same order as the electric part and thus,
for $d=4$, to have non-vanishing NUT charge. Definitive results
are demonstrated when the boundary is either a cylindrical or a
hyperbolic (i.e., de Sitter space) representation of spacelike
infinity ($i^0$), and partial results are provided for more
general representations of $i^0$. For the cylindrical or
hyperbolic representations of $i^0$, similar results are also
shown to hold for both a counter-term proportional to the
square-root of the boundary Ricci scalar and for a more
complicated counter-term suggested previously by Kraus, Larsen,
and Siebelink. Finally, we show that such actions lead, via a
straightforward computation, to conserved quantities at spacelike
infinity which agree with, but are more general than, the usual
(e.g., ADM) results.}

\date{September 2005}

\keywords{Asymptotic flatness, gravitational action}

\preprint{hep-th/0511096}

\begin{document}


\section{Introduction}
\label{intro}

In covariant approaches to quantum mechanics, the action $S$
provides a fundamental link between classical and quantum
treatments. Broadly speaking, classical mechanics is recovered
through the semi-classical approximation, in which the path
integral is dominated by stationary points of the action. Here we
note that, in order to dominate the path integral, the action must
be stationary under the \textit{full} class of variations
corresponding to the space of paths over which the integral is
performed. Thus, one must consider \textit{all} variations which
preserve any boundary conditions and not just, say, variations of
compact support. In particular, requiring the action to be
stationary should yield precisely the classical equations of
motion, with all boundary terms in the associated computation
vanishing on any allowed variation.

Even ignoring the low differentiability of paths in the support of the
measure and restricting the discussion to smooth paths, this requirement can be rather subtle.
We are interested here in the case of asymptotically flat gravity.  Thus, we begin with the familiar covariant action given by the
Einstein-Hilbert ``bulk'' term with Gibbons-Hawking boundary term,
\begin{equation}  \label{EH+GH}
S_{EH+GH} = - \frac{1}{16\pi G} \int_{\mathcal{M}} \sqrt{-g} R - \frac{1}{%
8\pi G} \int_{\partial \mathcal{M}} \sqrt{-h} K.
\end{equation}
This action does {\it not} satisfy the above requirement under all asymptotically flat variations. Indeed, as is well-known,
on the space of classical solutions (\ref{EH+GH}) satisfies
\begin{equation}
\delta S_{EH+GH} = \frac{1}{16\pi G} \int_{\partial \mathcal{M}} \sqrt{-h}
\pi^{ij} \delta h_{ij},  \label{varEHGH}
\end{equation}
where $\pi^{ij} = K^{ij} - Kh^{ij}$ and $h_{ij}$
is the induced metric on a timelike boundary $\partial \mathcal{M}$. Here we
have, for simplicity, neglected additional boundary terms associated with
past and future boundaries, and we will continue to neglect such terms
throughout this work. The metric $h_{ij}$ and its inverse are used to raise
and lower indices $i,j,k,l,m...$ associated with the boundary spacetime.

The reader will readily check that, when the boundary $\partial \mathcal{M}$ is
taken to spatial infinity, the right-hand side of (\ref{varEHGH}) does
\textit{not} vanish under standard definitions of asymptotic
 flatness, e.g., \cite{WaldBook, AshtekarHansen, AshtekarInf,ABR, AshtekarRomano}. 
Instead,
for standard choices\footnote{
Recall that the boundary spacetime $(\partial M, h)$ is not
uniquely defined by the bulk spacetime $(\mathcal{M}, g)$.
Instead, it also depends on the choice of limiting procedure used
to define the Gibbons-Hawking term. This issue will be discussed
in detail in section \ref{prelim} below.  The divergence of
(\ref{varEHGH}) is linear for what will be called ``hyperbolic''
temporal cut-offs in section \ref{prelim}, and (\ref{varEHGH})
approaches a constant under what will be called ``cylindrical''
temporal cut-offs.} of $%
\partial M$, the variation (\ref{varEHGH}) generically either diverges
linearly or approaches a non-zero constant as $\partial M$ is taken to
spatial infinity.

In contrast, the analogous stationarity requirement has been well-studied
within the 3+1 framework. Indeed, Regge and Teietelboim \cite{RT}
showed that requiring the Hamiltonian to be stationary under
all asymptotically flat variations leads directly to the ADM
boundary term \cite{ADM1,ADM2,ADM3}, and thus to the usual ADM
definitions of energy, momentum, and angular momentum at spacelike infinity.
Similarly, in the Palatini formalism, fixing a gauge at infinity for local Lorentz transformations allows a boundary term \cite{eeA,ABL}  which yields a well-defined variational principle for asymptotically flat spacetimes.
What we seek is an analogous covariant boundary term for the Einstein-Hilbert action\footnote{Both the Palatini and canonical variational principles mentioned above are valid only for what are called ``cylindrical'' temporal cut-offs below.  Thus, we also seek extensions to the ``hyperbolic" temporal cut-offs.}. We will find that
it can be
specified by a local, covariant term analogous to the
counter-terms used to regulate the gravitational action of
asymptotically anti-de Sitter
spacetimes, see e.g. \cite {skenderis}-\cite{KS9} .

Within the covariant framework, the ``reference background
approach''
provides a step toward our goal.
Here one adds to $S_{EH+GH}$ an additional term
\begin{equation}
S_{Ref}=\frac{1}{8\pi G}\int_{\partial \mathcal{M}}\sqrt{-h}K_{Ref},
\label{HH}
\end{equation}%
where $K_{Ref}$ is the trace of the extrinsic curvature of the boundary $(\partial \mathcal{M},h)$ when this boundary spacetime is detached from $\mathcal{M}$ and isometrically embedded in Minkowski
space (or, more generally, in another reference background, see e.g. \cite{BY,BCM,HawHor}.)
As with the Gibbons-Hawking term, the term $S_{Ref}$ is to be defined
through an appropriate limiting procedure.

The term (\ref{HH}) and
its generalizations were originally suggested (see, e.g.
\cite{HawHor}) in order to make the action finite on the space of
solutions,  but variations of this term are not typically
addressed in the literature.  Indeed, before one can vary this term one must address the
existence and uniqueness of embeddings of $(\partial {\cal M},h)$
into $({\cal M}^{Ref}, g^{Ref})$.  Note that, in order to vary
$S_{Ref}$, we will need such an embedding not only for some
particular boundary spacetime, but in fact for an open set of
boundary spacetimes associated with arbitrary small variations.

In $d=3$ spacetime dimensions, the desired embeddings may plausibly exist
for suitably general choices of $(\partial \mathcal{M},h)$ near
spatial infinity of asymptotically flat spacetimes.  For example,
in the Euclidean signature, Weyl's embedding theorem states that a
2-manifold ($\partial M, h$) can be embedded in Euclidean
${\mathbb R}^3$ if the scalar curvature of $h$ is everywhere
positive. (It also states that the resulting extrinsic curvature
can be chosen to be positive definite.)

However, such an embedding generically fails in higher dimensions:
in particular, given any boundary spacetime which can be so
embedded, there are spacetimes arbitrarily nearby which cannot. To
see this, simply note that any co-dimension one embedding can be
specified locally by a single relation among the coordinates of
the embedding space; i.e., by a single function on $\partial M$.
In contrast, when the dimension of $\partial M$ is greater than 2,
the metric $h$ (after gauge-fixing) contains more than a single
degree of freedom\footnote{This argument was suggested to us by
Jan de Boer.}. Thus, an open set of such embeddings can exist only
for $d\leq 3$, and variations of $S_{Ref}$ (\ref{HH}) are
ill-defined in $d \ge 4$ dimensions.

Nevertheless, in this work we show that, for asymptotically flat vacuum spacetimes with $%
d\geq 4$, one may build the desired action by simply replacing
$K_{Ref}$ in (\ref{HH}) with the trace of the tensor
$\hat{K}_{ij}$ (i.e., $\hat{K}=h^{ij}\hat{K}_{ij}$) implicitly
defined by solving the equation
\begin{equation}
\mathcal{R}_{ik}=\hat{K}_{ik}\hat{K}-\hat{K}_{i}^{m}\hat{K}_{mk},
\label{GCM}
\end{equation}%
where $\mathcal{R}_{ij}$ is the Ricci tensor of $h_{ij}$ induced
on $\partial {\cal M}$.  Although the definition is implicit,
our counter-term is clearly a local algebraic function of the boundary
metric $h_{ij}$ and its Ricci tensor ${\cal R}_{ij}$.  Adding such
a counter-term to the Einstein-Hilbert action with Gibbons-Hawking
boundary term results in an action which is finite on
asymptotically flat solutions, and which has well-defined
asymptotically flat variations which vanish on solutions.

The motivation for using equation (\ref{GCM}) to define $\hat K$ arises from considering
the Gauss-Codazzi relation
(see, e.g. \cite{WaldBook}) for a surface with spacelike normal,
\begin{equation}
\mathcal{R}_{ijkl}={R}^{Ref}_{ijkl}+K_{ik}K_{jl}-K_{jk}K_{il}.
\label{GCunTr}
\end{equation}%
In particular, an equation of the form (\ref{GCunTr}) would hold
if the boundary spacetime $(\partial \mathcal{M},h)$ had indeed
been embedded in a reference background $({\cal
M}^{Ref},g^{Ref})$. Here $\mathcal{R}_{ijkl}$ is the Riemann
tensor (formed from $h$) on $\partial \mathcal{M}$ and
$R^{Ref}_{ijkl}$ is the (bulk) Riemann tensor of $({\cal M}^{Ref},
g^{Ref})$ pulled back to $\partial \mathcal{M}$.  In the case
where the reference background is Minkowski space, we have
$R^{Ref}_{ijkl}=0$. Note that for $d-1> 3$ the full Gauss-Codazzi
relation (\ref{GCunTr}) has too many components to generically
admit a solution for the extrinsic curvature, $K_{ij}$.  However,
taking the trace of (\ref{GCM}) by contracting with $h^{il}$
provides a symmetric tensor equation (\ref{GCM}), which may then
be solved for $K_{ij}$ (also a symmetric tensor) within open sets
in the space of possible $\mathcal{R}_{ij}$. In particular, we
will see that solutions always exist for suitable definitions of
$\partial M$ near spacelike infinity of an asymptotically flat
spacetime; i.e., for all spacetimes over which the action should
be varied. Note that, although the definition of $\hat K$ is
somewhat implicit, $\hat K$ is nevertheless
a local algebraic function of the boundary metric $%
h_{ij}$ and its Ricci curvature $\mathcal{R}_{ij}$.

The plan of our paper is as follows: Section \ref{ct} begins with
various preliminaries and definitions (section \ref{prelim}),
including our definition of asymptotically flat spacetimes.  The
reader will note that our notion of asymptotic flatness is less
restrictive than that of many standard references
\cite{AshtekarHansen,ABR,AshtekarRomano,BS,B}, as we allow the
electric and magnetic parts of the Weyl tensor to be of the same
order; in particular, for $d=4$  we allow non-vanishing NUT
charge.

Section \ref{ct} contains our main results.  In particular, we
show that the action
\begin{equation}
\label{Srenorm}
S_{renorm}=S_{EH+GH}+S_{new \ CT},
\end{equation}%
with
\begin{equation}
S_{new \ CT}=\frac{1}{8\pi G}\int_{\partial
\mathcal{M}}\sqrt{-h}\hat{K}, \label{Sct}
\end{equation}%
where $\hat K$ is defined via the solution of (\ref{GCM}),
leads to a finite action on asymptotically flat solutions (section \ref{newct}) and that (section %
\ref{varSrenorm}) the action is stationary under all
asymptotically flat
variations for common ``cylindrical'' and ``hyperbolic'' choices of $({%
\partial M},h)$.
For a large class of more general boundaries, we show
that (\ref{Srenorm}) is again finite on asymptotically flat solutions, but we
reach no conclusion regarding its variations.

We then turn in section \ref{other} to two other proposed counter-terms \cite%
{Mann,KLS} for asymptotically flat spacetimes.  We show, for
cylindrical and hyperbolic choices of $({\partial M},h)$, that for $d \ge 5$
such counter-terms again define successful covariant variational
principles for asymptotically flat spacetimes with cylindrical and
hyperbolic boundaries.  However, for at least one of these additional two counter-terms, the corresponding result in $d=4$ holds only for the cylindrical boundaries.

In section \ref{CQ}, we discuss conserved quantities (energy,
angular momentum, etc.) constructed from our renormalized actions.
We show that, when they provide valid variational principles, all
of the above actions lead via the algorithm of \cite{HIM2} to
(finite) conserved quantities at spatial infinity for each
asymptotic symmetry.  Our conserved quantities agree with the
usual definitions
\cite{AshtekarHansen,ABR,AshtekarRomano,RT,ADM1,ADM2,ADM3} of
energy, momentum, etc., but also generalize such definitions to
allow, e.g., non-vanishing NUT charge in
four-dimensions\footnote{An independent construction, based on canonical methods,
of conserved
quantities for $d=4$ in the presence of NUT charge will appear soon \cite{HTnew}.}.  We also
demonstrate that these quantities are related to a boundary stress
tensor of the sort used in \cite{DAR,Mann2} (and related to that
of the anti-de Sitter context \cite{skenderis,kraus}). Finally, we
close with some discussion in section \ref{disc}.

\section{Preliminaries}

\label{prelim}

This section presents various definitions and fixes notation for
the rest of the paper.  We begin with the definition of asymptotic
flatness used in this work.

Consider a $d\ge 4$ dimensional spacetime for which the line
element admits an expansion of the form\footnote{The inclusion of NUT-charge requires some changes in the global structure, but these changes have little effect on the arguments below.  Instead of presenting the details here, we confine ourselves to brief comments on the NUT case in the relevant places below.}
\begin{equation}  \label{AFdef}
ds^2 = \left( 1+ \frac{2 \sigma}{\rho^{d-3}} + \mathcal{O}(\rho^{-(d-2)})%
\right) d\rho^2 + \rho^2 \left( h^0_{ij} + \frac{h^1_{ij}}{\rho^{d-3}}
+ \mathcal{O}(\rho^{-(d-2)}) \right) d\eta^i d\eta^j,
\end{equation}
for large positive $\rho$. Here, $h^0_{ij}$ and $\eta^i$ are a metric and
the associated coordinates on the unit $(d-2,1)$ hyperboloid ${\cal H}^{d-1}$
(i.e., on $d-1$ dimensional de Sitter space) and $\sigma, h^1_{ij}$ are
respectively a smooth
function and a smooth tensor field on ${\cal H}^{d-1}$.
Thus, $\rho$ is the ``radial'' function
associated with some asymptotically Minkowski coordinates $x^a$ through $%
\rho^2 = \eta_{ab} x^a x^b.$ In (\ref{AFdef}), the symbols $\mathcal{O}%
(\rho^{-(d-2)})$ refer to terms that fall-off at least as fast as $%
\rho^{-(d-2)}$ as one approaches \textit{spacelike} infinity, i.e., $\rho
\rightarrow +\infty$ with fixed $\eta$. For the purposes of this paper, we
shall take (\ref{AFdef}) to define the class of asymptotically flat
spacetimes. Here we follow a coordinate-based approach, but a more geometric treatment of this condition can also be given, generalizing to higher dimensions either the treatment of
Ashtekar and Hansen \cite{AshtekarHansen} or that of Ashtekar and Romano %
\cite{AshtekarRomano}.

Note that, for $d=4$, any metric which is asymptotically flat by
the criteria of any of
\cite{WaldBook,AshtekarHansen,ABR,AshtekarRomano} also satisfies
(\ref{AFdef}).    In $d\ge 5$ dimensions, our definition is more
restrictive than that of \cite{skenAF}, which for $d \ge 5$
allows additional terms of order $\rho^{-k}$ for $d-4 \ge k \ge 1$
relative to the leading terms.  However, our definition is at
least as general as the definition which would result by applying
the methods of \cite{RT}; i.e., by considering the action of the
Poincare' group on the Schwarzschild spacetime. We also note that,
because Minkowski space itself solves the equations of motion (the
Einstein equations), it is clear that (\ref{AFdef}) is consistent
with the dynamics of the system.

Consider now the action $S_{renorm}$, including in particular the
``counter-term'' $S_{new \ CT}$ of (\ref{Sct}). We wish to show that $%
S_{renorm} $ is finite on asymptotically flat solutions (i.e., on Ricci flat
spacetimes satisfying (\ref{AFdef})), and that it is stationary about any
such solution under variations preserving (\ref{AFdef}).  In general, we will allow any variation compatible with (\ref{AFdef}).  The one exception will occur in the case of $d=4$ spacetime dimensions, and will be mentioned shortly (see equation \ref{varh1}).

To derive such results, we must carefully specify the form of the boundary
spacetime $(\partial M, h)$. As with any discussion of the more familiar
Gibbons-Hawking term in asymptotically flat spacetimes, the term `boundary
spacetime' is an abuse of language which in fact refers to a one-parameter
family $(\partial \mathcal{M}_\Omega, h_\Omega)$ of boundaries of regions $%
\mathcal{M}_\Omega \subset \mathcal{M}$. Here we take $\mathcal{M}_\Omega$
to be an increasing family (i.e., $\mathcal{M}_\Omega \supset \mathcal{M}%
_{\Omega^{\prime}}$ whenever $\Omega > \Omega^{\prime}$) converging to $%
\mathcal{M}$ (that is, such that $\cup_\Omega \mathcal{M}_\Omega = \mathcal{M%
}$). Any such family represents a particular way of `cutting off' the
spacetime $\mathcal{M}$ and then removing this cut-off as $\Omega
\rightarrow \infty$. Thus, expressions such as (\ref{EH+GH}), (\ref{HH}),
and (\ref{Sct}) are to be understood as the $\Omega \rightarrow \infty$
limits of families of functionals $S_\Omega$, in which $(\mathcal{M}, {%
\partial \mathcal{M}}, {\ h})$ are replaced by $(\mathcal{M}_\Omega, {%
\partial \mathcal{M}}_\Omega, {h}_\Omega)$. We will take this cut-off to be
specified by some given function $\Omega$ on $\mathcal{M}$ such that $\Omega
\rightarrow \infty$ at spatial infinity. We define $\mathcal{M}_{\Omega_0}$
to be the region of $\mathcal{M}$ in which $\Omega < \Omega_0$, so that $%
(\partial \mathcal{M}_{\Omega_0}, h_{\Omega_0})$
is the hypersurface where $\Omega = \Omega_0$.

Two particular classes of cut-off $\Omega$ will be of use below.
The first is the class of ``hyperbolic cut-offs,'' in which
$\Omega$ is taken to be some function of the form:
\begin{equation}  \label{hypcut}
\Omega^{hyp} = \rho + O(\rho^{0}),
\end{equation}
where the fall-off condition on $\Omega^{hyp}$ is chosen so that the metric induced
by (\ref{AFdef}) on any surface $\Omega^{hyp} = constant$ takes the form
\begin{equation}  \label{HypInduce}
h_{ij} = \rho^2 \left( h^0_{ij} + \frac{h^1_{ij} }{\rho^{d-3}}+
\mathcal{O}(\rho^{-(d-2)}) \right),
\end{equation}
where $h^0_{ij}$ and $h^1_{ij}$ were defined previously in equation (\ref{AFdef}).
Choosing such a cut-off leads to a hyperbolic representation of spacelike
infinity analogous to the construction of Ashtekar and Romano \cite%
{AshtekarRomano}, though we will not pursue all details of the
geometric structure here. Another useful class of cut-offs is the
``cylindrical cut-off''
\begin{equation}  \label{cylcut}
\Omega^{cyl} = r + O(\rho^{0}).
\end{equation}
In (\ref{cylcut}), the coordinate $r$ is defined by $r^2 = \rho^2 + t^2$ and
$t$ is an asymptotically Minkowski time coordinate. More precisely, we may
define $t$ through the requirement that the metric (\ref{AFdef}) takes the
form
\begin{equation}  \label{cyldef}
ds^2 = - \left( 1+ \mathcal{O}(\rho^{-(d-3)})\right) dt^2 + \left( 1+
\mathcal{O}(\rho^{-(d-3)})\right) dr^2 + r^2 \left( \mu_{IJ} +\mathcal{O}%
(\rho^{-(d-3)}) \right) d\theta^I d\theta^J,
\end{equation}
where $\mu_{IJ}, \theta^I$ are the metric and coordinates on the unit $%
(d-2)$-sphere. As implied by (\ref{hypcut}) and (\ref{cylcut}), the action $%
S_{renorm}$ will depend only on the asymptotic form of $\Omega$, which we
will take to represent a fixed auxiliary structure.

A further subtlety is related to the way in which the spacetime
$\mathcal{M}$ is cut off in time. In the Lorentz-signature context
(on which we focus), one is typically interested in the region of
spacetime between two Cauchy
surfaces (say, $\Sigma_-$ and $\Sigma_+$), and we will have in mind that $%
\mathcal{M}$ represents such a region. However, in the asymptotically flat
setting, two rather different such situations may be natural, depending on the
physical context. Consider first the special case in which $\Sigma_+$ and $%
\Sigma_-$ are related by an asymptotic translation. Then the volume of $%
\partial \mathcal{M}_\Omega$ grows as $\rho^{d-2}$ in the limit $\Omega \rightarrow
\infty$.   We refer to spacetimes ${\cal M}$ having past and future boundaries related in this way as
corresponding to ``a cylindrical temporal cut-off $\mathcal{T}^{cyl}$."

However, it is also natural to consider a more general case in
which $\Sigma_+$ and $\Sigma_-$ are allowed to asymptote to any
Cauchy surfaces $C_+$ and $C_-$ of the hyperboloid ${\cal
H}^{d-1}$; that is, we allow $\Sigma_\pm$ to be defined by any
equations of the form
\begin{equation}
0 =  f_\pm (\eta) + {\cal O}(\rho^{-1}),
\end{equation}
for smooth functions $f_\pm$ on ${\cal H}^{d-1}$.  One may think
of such surfaces $\Sigma_-,\Sigma_+$ as being locally boosted
relative to each other at infinity.   We refer to spacetimes
${\cal M}$ having this latter sort of past and future boundaries
as corresponding to ``a hyperbolic temporal cut-off
$\mathcal{T}^{hyp}$." Note that when ${\cal M}$ is defined by such
past and future boundaries the volume of $\partial
\mathcal{M}_\Omega$ grows as $\rho^{d-1}$.

Since the  volume of $\partial {\cal M}$ grows as one power of $\rho$
faster in the case of hyperbolic temporal cut-off (${\cal
T}^{hyp}$) than in the case of cylindrical temporal cut-off
($\mathcal{T}^{cyl}$), it is clear that the choice of temporal
cut-off can affect conclusions about our action $S_{renorm}$. One
consequence is that, in the particular case of $d=4$ and
for the hyperbolic temporal cut-off, it will be necessary to
restrict variations of $h^1_{ij}$ to be of the form
\begin{equation}
\label{varh1} \delta h^1_{ij} = \alpha h^0_{ij},
\end{equation}
for $\alpha$ a smooth function on ${\cal H}^{3}$.  The physics of this
restriction will be discussed further in section \ref{varSrenorm} and in
appendix \ref{restrict}.
Here we simply note that this restriction is analogous to
a condition imposed in \cite{ABR} in order to arrive at a well-defined covariant phase space formalism.

In physical applications, it is natural to use a cylindrical
temporal cut-off $\mathcal{T}^{cyl}$ in conjunction with the
cylindrical spatial cut-off $\Omega^{cyl}$; for example when
studying evolution between two Cauchy surfaces related by time
translations or, in the Euclidean context, with periodic Euclidean
time (i.e., at finite temperature). Similarly, it seems natural to
use a hyperbolic temporal cut-off $\mathcal{T}^{hyp}$ in
conjunction with a hyperbolic spatial cut-off $\Omega^{hyp}$ (as
in, for example, the covariant phase space treatment of Ashtekar,
Bombelli, and Reula \cite{ABR}). However, in principle one may make
independent choices of spatial and temporal cut-off. We find it
interesting to do so below in order to probe the possibility of
generalizing our results to a more general spatial cut-offs. In
particular, one would like to generalize the results below to
arbitrary spatial cut-offs of the form $\Omega = \omega
\Omega^{hyp}$, with $\omega$ a smooth non-vanishing function on
the unit hyperboloid.  Our results indicate that this may be possible, at least in the case of cylindrical
temporal cut-off ${\cal T}^{cyl}$.

\section{Gravitational counter-terms}

\label{ct} In this section we study the new counter-term
(\ref{Sct}) for
asymptotically flat spacetimes, as well as those counter-terms suggested previously in \cite%
{Mann,KLS}. Our main results for the new counter-term are presented in sections
\ref{newct} and \ref{varSrenorm}. Similar results are derived for
the counter-terms of \cite{Mann,KLS} in section \ref{other}.

\subsection{The on-shell action is finite}

\label{newct}

We now consider our new counter-term and show that the action is
finite on asymptotically flat (see \ref{AFdef})
solutions of the equations of motion. For a cylindrical temporal cut-off $%
\mathcal{T}^{cyl}$, this is a straightforward exercise in power counting for
(almost) any spatial cut-off $\Omega =\omega \Omega ^{hyp}$ with $\omega $ a
smooth non-vanishing function on the hyperboloid and, in particular, for $%
\Omega =\Omega ^{hyp}$ or $\Omega =\Omega ^{cyl}$. To see this,
simply note that, since our spacetimes are Ricci flat, the
Einstein-Hilbert term vanishes on solutions.  As a result,
only the boundary terms contribute to the action:
\begin{equation}
S_{renorm}=-\frac{1}{8\pi G}\int \sqrt{-h}(K-\hat{K}).  \label{onshell}
\end{equation}%

To first order, the difference $\Delta K_{ij}:=K_{ij}-\hat{K}_{ij}$ can
be found by linearizing the trace of the Gauss-Codazzi relations (\ref%
{GCunTr}). Furthermore, since both $K_{ij}$ and $\hat{K}_{ij}$ satisfy such (traced) relations
with the same metric $h_{ij}$ and Ricci tensor $\mathcal{R}_{ij}$,
the change $\Delta K_{ij}$ is sourced entirely by the bulk Riemann tensor:
\begin{equation}
h^{kl}{R}_{ikjl}=-\Delta  {K}_{kl}\left( h^{kl}\hat{K}_{ij}+\delta
_{i}^{k}\delta _{j}^{l}\hat{K}-\delta _{i}^{k}\hat{K}_{j}^{l}-\delta _{j}^{k}%
\hat{K}_{i}^{l}\right).  \label{linGC}
\end{equation}
Note that $h^{jk}$ is the inverse of $h_{ij}$, and that we will
always raise and lower indices $i,j,k,l,...$ with $h^{ij}$ and
$h_{ij}$ (as opposed to using $(h^0)^{ij}$ and $h^0_{ij}$).

Now, the operator
\begin{equation}
\label{L}
L^{kl}_{ij} =   h^{kl}\hat{K}_{ij}+\delta
_{i}^{k}\delta _{j}^{l}\hat{K}-\delta _{i}^{k}\hat{K}_{j}^{l}-\delta _{j}^{k}%
\hat{K}_{i}^{l},
\end{equation}
acting on $\Delta K_{kl}$ in (\ref{linGC}) is generically invertible and of the same order in $\rho$ as $K$; namely, ${\cal O}(\rho^{-1})$.  Thus, $L^{kl}_{ij}$ will generically
have an inverse of order $\mathcal{O}(\rho )$; we will see this explicitly for $\Omega^{cyl}$ and $\Omega^{hyp}$ in section \ref{varSrenorm}. Since $%
h^{mn}R_{kmln}=\mathcal{O}(\rho ^{-(d-3)})$, we have
$\Delta K_{ij}=\mathcal{O}(\rho
^{-(d-4)})$, and higher corrections to $\Delta K_{ij}$ are sub-leading. As a
result, $\Delta K=K-\hat{K}=\mathcal{O}(\rho ^{-(d-2)})$ and, with a temporal
cut-off $\mathcal{T}^{cyl}$, we have $\int_{\partial {\cal M}} \sqrt{-h}\Delta K=\mathcal{O}(1)$.

Let us now address the case of hyperbolic temporal cut-off
$\mathcal{T}^{hyp}$.  Because the volume element on $\partial {\cal M}$ is
larger by a power of $\rho$,  a more careful
analysis is required to obtain useful results in this case and, in
the end, we will obtain such results only for hyperbolic spatial
cut-off $\Omega^{hyp}$.  Our starting point is the observation that, for $\Omega^{hyp}$,
substituting
$\hat{K}_{ij}=\frac{1}{\rho }h_{ij}+\mathcal{O}(\rho ^{-{(d-4)}})$ into (\ref{linGC}),
yields
\begin{equation}
\left( \frac{d-3}{\rho }+\mathcal{O}(\rho ^{-{(d-2)}})\right) \Delta
K_{ik}=-R_{ijkl}h^{jl}+\frac{1}{2(d-2)}h_{ij}h^{mn}R_{mknl}h^{kl}.
\label{step2}
\end{equation}%

Now, recall that we are interested in vacuum \textit{solutions};
i.e., Ricci-flat
metrics\footnote{More generally, one may consider matter sources with $T_{ab}=%
\mathcal{O}(\rho ^{-d})$ in terms of asymptotically Minkowski Cartesian coordinates.
Since the Weyl tensor in such coordinates is typically of order $\rho ^{-(d-1)}$, the Riemann
and Weyl tensors agree to leading order, so that one may proceed similarly in
such cases.}. As a result, we may replace $R_{ijkl}$ with the bulk Weyl tensor, $%
C_{ijkl}$. Recall that the Weyl tensor is traceless: $g^{ac}C_{abcd}=0=g^{bd}C_{abcd},$
where $g_{ab}$ is the metric on $\mathcal{M}$ and indices $a,b,c...$ will
denote coordinates on $\mathcal{M}$. Introducing the unit normal $N^{a}$ to
the boundary, we may define
\begin{equation}
h_{ab}=g_{ab}-N_{a}N_{b},
\end{equation}%
which pulls back to the metric $h_{ij}$ on the boundary. Furthermore, $%
h_{b}^{a}$ is a projector onto directions tangent to ${\partial \mathcal{M}}$%
. It follows that the electric part of the Weyl tensor (as in \cite%
{AshtekarHansen,AshtekarInf,AshtekarRomano}) is
\begin{equation}
E_{ac}:=C_{abcd}N^{b}N^{d}=-C_{abcd}h^{bd}
\end{equation}%
and that, due to the antisymmetry and traceless properties of the Weyl
tensor, we have $h^{ac}E_{ac}=0$. We may then pull this expression back to
the boundary to find the familiar result $h^{ij}E_{ij}=0$. Thus, (\ref{step2}%
) yields
\begin{equation}
K_{ij}-\hat{K}_{ij}=\frac{\rho }{d-3}E_{ij}+\mathcal{O}(\rho ^{-(d-3)}).
\label{step3}
\end{equation}%
But $E_{ij}$ is traceless, so the contribution to $K-\hat{K}$ from
$E_{ij}$ vanishes and $K-\hat{K}=\mathcal{O}(\rho ^{-(d-1)})$.
Thus, for a hyperbolic temporal cut-off ${\cal T}^{hyp}$, the
integrand in (\ref{onshell}) is of order $\rho ^{0}$.  We conclude
that,  neglecting past and future boundary terms, $S_{renorm}$
takes finite values on asymptotically flat solutions with
$(\partial \mathcal{M})$ defined by $\Omega ^{hyp}$.

 We have shown above that $S_{renorm}$ is finite for either i) cylindrical temporal cut-offs (${\cal T}^{cyl}$) and any spatial cut-off $\Omega=\omega \Omega^{hyp}$, where $\omega$ is a smooth function on the hyperboloid, or ii) hyperbolic temporal cut-off (${\cal T}^{hyp}$) with hyperbolic spatial cut-off ($\Omega^{hyp}$.  However, we have not yet addressed variations of such actions, nor have we discussed whether the numerical value of each action is sensitive to the particular spatial cut-off $\Omega$, say, within the class $\Omega^{hyp}$ (\ref{hypcut}).  Both of these issues will be addressed in subsection \ref{varSrenorm} below.

\subsection{Variations of the action}

\label{varSrenorm}

Having shown that $S_{renorm}$ is finite on solutions, we now consider its
first variations. We wish to show that such variations vanish about any
asymptotically flat solution using either spatial cut-off, $\Omega^{cyl}$ or
$\Omega^{hyp}$, and using either temporal cut-off ${\cal T}^{cyl}$ or ${\cal T}^{hyp}$.  For the particular case of $d=4$ and ${\cal T}^{hyp}$, we will require the variations to satsify the extra condition (\ref{varh1}).

Since the variation of the Einstein-Hilbert action with Gibbons Hawking term is given by
(\ref{varEHGH}), our task is essentially to consider the variation of our new counter-term $S_{new  \ CT}$, which we compute as follows:
\begin{equation}
\delta S_{new \ CT}=\frac{1}{8\pi G}\int_{\partial M}\delta (\sqrt{-h}\hat{K})=%
\frac{1}{8\pi G}\int_{\partial M}\sqrt{-h}\left( \frac{1}{2}\hat{K}%
h^{ij}\delta h_{ij}+\hat{K}_{ij}\delta h^{ij}+h^{ij}\delta \hat{K}%
_{ij}\right) .  \label{v1}
\end{equation}%
Now, expanding the definition (\ref{GCM}) of $\hat{K}_{ij}$ to first order,
we find
\begin{eqnarray}
\delta \mathcal{R}_{ij} &=&\delta \hat{K}_{kl}\left( h^{kl}\hat{K}%
_{ij}+\delta _{i}^{k}\delta _{j}^{l}\hat{K}-\delta _{i}^{k}\hat{K}%
_{j}^{l}-\delta _{j}^{k}\hat{K}_{i}^{l}\right) \nonumber \\
&+&(\hat{K}_{ij}\hat{K}_{mn}-\hat{K}_{im}\hat{K}_{nj})\delta h^{mn} \label{toSolve}.
\end{eqnarray}
For the case where $h^1_{ij}, \sigma$ and the higher corrections to the metric vanish,
(\ref{GCM}) is easy to solve.  This solution is
\begin{equation}
\label{K}   \ \ \ \hat K_{ij}  = \Bigg\{ {{
\frac{1}{\rho} h_{ij}  = \rho (h^0)_{ij} \hspace{3.9cm}  {\rm for } \ \Omega^{hyp }} \atop { r
\mu_{ij}
  \ \ \ \hspace{5.2cm}
 {\rm for } \ \Omega^{cyl},
 }}
\end{equation}
where $\mu_{ij}$ is the pull-back of $\mu_{IJ}$ (the unit round metric on $S^{d-2}$)  to
$S^{d-2} \times \mathbb{R}$.
More generally, we can solve perturbatively around this solution.

By contracting (\ref{toSolve})  with $h^{ij}$ and $\mu ^{ij}$, it
is straightforward to show that the trace of $\delta \hat{K}_{ij}$
satisfies
\begin{equation}
\label{SolvePert} \delta \hat K_{ij} h^{ij} = \Bigg\{ {{-
\frac{1}{2} \left[ \hat K_{ij} \delta h^{ij} - \frac{\rho}{(d-2)}
h^{ij} \delta {\cal R}_{ij} \right] + {\cal O}(\rho^{-(2d-5)})   \hspace{4.05cm}  {\rm for } \
\Omega^{hyp }} \atop { -\frac{1}{2} \hat K_{ij} \delta h^{ij} +
\frac{1}{(2d-6)\hat K} \left[ (d-4) h^{ij} \delta {\cal R}_{ij} +
\frac{2}{r^2} \mu^{ij} \delta {\cal R}_{ij} \right]  + {\cal O}(\rho^{-(2d-5)})  \ \ \
 {\rm for } \ \Omega^{cyl},}}
\end{equation}
and in fact to iteratively solve for the full $\delta K^{ij}$. As
an aside we note that, using the inverse function theorem and
compactness of $\partial \mathcal{M}$ (due to the temporal
cut-off), we may conclude that, given any asymptotically flat
$({\cal M}, g)$, solutions for $\hat{K}_{ij}$ exist for
sufficiently large $\Omega $.

Now, recall \cite{WaldBook} that $\delta \mathcal{R}_{ij}$ can be
expressed in the form
\begin{equation}  \label{dR}
\delta \mathcal{R}_{ij} = - \frac{1}{2} h^{kl} D_i D_j \delta h_{kl} - \frac{%
1}{2} h^{kl} D_k D_l \delta h_{ij} + h^{kl} D_k D_{(i} \delta
h_{j)l},
\end{equation}
where $D_i$ is the (torsion-free) covariant derivative on ${\cal
H}^{d-1}$ compatible with $h_{ij}$.  Thus, when $\delta
\mathcal{R}_{ij}$ is contracted with any covariantly constant
tensor  (e.g., $(h^0)^{ij}$
or, on the cylinder, $\mu^{ij}$), the
result is a total derivative. Note that because
we are expanding about homogeneous spaces, $\hat K$ is constant over the
space, so that $\frac{1}{\hat{K}}h^{ij}\delta R_{ij}$ yields a total divergence in
the
cases of interest.

 Inserting the remaining term
$-\frac{1}{2}\hat K_{ij} \delta h^{ij}$ from (\ref{SolvePert})
into (\ref{v1}), up to boundary terms at the past and future
boundaries and when the equations of motion hold we find
\begin{equation}  \label{v2}
\delta S_{renorm} = \frac{1}{16\pi G} \int_{\partial M} \sqrt{-h} (\pi^{ij}
- \hat \pi^{ij}) \delta h_{ij}  + {\cal O} (\rho^{-(d-4+c)}),
\end{equation}
where $\hat \pi^{ij} = \hat K^{ij} - \hat K h^{ij}$ and we have used that $%
A^{ij} \delta h_{ij} = - A_{ij} \delta h^{ij}$ for any tensor $A_{ij}$. The constant $c$ in the exponent of $\rho$ in the error term takes the value $c=1$ for cylindrical temporal cut-offs (${\cal T}^{cyl}$) and takes the value $c=0$ for hyperbolic temporal cut-offs (${\cal T}^{hyp}$).  Thus, the error term contributes only for $d=4$ with hyperbolic temporal cut-off  (${\cal T}^{hyp}$).
Note
that $\hat \pi^{ij}$ has the same form as the momentum $\pi^{ij}$ conjugate to $%
h_{ij}$, except that it is built from $\hat K^{ij}$ instead of from the actual
extrinsic curvature $K^{ij}$.

For some cases, one may show that (\ref{v2}) vanishes on asymptotically flat solutions by simply
counting powers of $\rho $. We note from (\ref{step3}), (and the
analogous result for $\Omega^{cyl}$)
that $K^{ij}-\hat{K}^{ij}$ is of order $\rho ^{-d}$. Since $\delta h_{ij}$
is of order $\rho ^{-(d-5)}$, the integral in (\ref{v2}) is of order $\rho
^{-(d-4)}$ for hyperbolic temporal cut-off $\mathcal{T}^{hyp}$ and is of
order $\rho ^{-(d-3)}$ for cylindrical temporal
cut-off $\mathcal{T}^{cyl}$. Thus, our action is stationary on
asymptotically flat solutions with either spatial cut-off ($\Omega ^{hyp}$
or $\Omega ^{cyl}$) for either $d\geq 5$ and hyperbolic temporal cut-off $%
\mathcal{T}^{hyp}$ or for $d\geq 4$ and cylindrical temporal cut-off $\mathcal{T}%
^{cyl}$.
Note that our argument for $\Omega^{cyl}$ generalizes
readily to more complicated cylindrical boundaries appropriate to infinitely long strings, branes, etc., whose metric to leading
order matches that of the standard embedding of $S^n \times
{\mathbb R}^{d-n-1} $ (with $n \ge 2$) into Minkowski
space\footnote{ Since we work in Lorentz signature, it is
convenient to abuse notation and to understand  $S^{n}$ for
$n=d-1$  to represent the Hyperboloid ${\cal H}^{d-1}$. Boundaries
of the form $S^n \times {\mathbb R}^{d-n-1} $ were of interest in
\cite{KLS}.}, as well as to other products of maximally symmetric
manifolds.

Let us now consider the case of $d=4$ with hyperbolic temporal
cut-off ${\cal T}^{hyp}$. The error term not written explicitly in
(\ref{v2}) must now be calculated to the next order.  This
calculation is outlined in appendix \ref{ErrorTerm}, where it is
shown that the leading contribution vanishes, so that this term is
in fact of order $\rho^{-1}$. Thus, we may concentrate on the
explicit term (involving $\pi^{ij}-\hat \pi^{ij}$) on the
right-hand-side of (\ref{v2}).

Now in $d=4$
spacetime dimensions one does not in fact expect a covariant quantum path
integral to integrate over all metrics of the form (\ref{AFdef}).
Instead, one expects the proper domain of integration to reflect
the classical covariant phase space. We note that in \cite{ABR} it
was necessary to restrict variations of $h^1_{ij}$ as in
(\ref{varh1}); i.e.,  to be of the form $\delta h^1_{ij} = \alpha
h^0_{ij}$ where $\alpha$ is a smooth function on the hyperboloid
${\cal H}^3$.  This was done in order to make the symplectic structure finite and to ensure that the
symplectic flux through spatial infinity vanishes.
Recall that finiteness of the symplectic structure is closely
related to the norm of perturbative particle states when one
quantizes the theory; this norm is just the symplectic product of
a positive frequency solution with its (negative frequency)
complex conjugate.  Thus, variations not compatible with keeping
the symplectic structure finite are properly viewed as a change of
boundary conditions, and not as a variation of histories within a
given physical system.  Similar comments apply to variations not
compatible with keeping symplectic flux from flowing outward
through spatial infinity.  For this reason, we are happy to adopt
the restriction (\ref{varh1}) here.

In fact, \cite{ABR} imposed  $h^1_{ij} = -2\sigma h^0_{ij}$, and
imposed as well as a number of other restrictions.  We shall have
no need for these additional restrictions.  However, since
\cite{ABR} justified the condition (\ref{varh1}) only in the
presence of these additional restrictions, it is legitimate to ask
whether we may use (\ref{varh1}) with more generality.  In
appendix \ref{restrict}, we argue that the answer is affirmative
by demonstrating that (\ref{varh1}) is compatible with the
equations of motion.

We will now show that,
with the restriction (\ref{varh1}),
 our action is stationary on asymptotically flat solutions when one chooses \textit{both} the spatial and temporal cut-offs to be hyperbolic ($\mathcal{T%
}^{hyp}$ and $\Omega^{hyp}$).  Note that we have
\begin{equation}
\delta h_{ij} = \frac{\alpha}{\rho^{d-5}} h^0_{ij} + \mathcal{O}%
(\rho^{-(d-4)}) = \frac{\alpha}{\rho^{d-3}} h_{ij} + \mathcal{O}%
(\rho^{-(d-4)}) .
\end{equation}
For this case, the leading order term in $(\pi^{ij} - \hat \pi^{ij}) \delta h_{ij}$
is proportional to the trace of $\pi^{ij} - \hat \pi^{ij}$.
 But, as shown in (\ref{step3}), this vanishes for hyperbolic
spatial cut-off $\Omega^{hyp}$. As a result, up to past and future
boundary terms, we
have
\begin{equation}  \label{v5}
\delta S_{renorm} = \frac{1}{16\pi G} \int_{\partial M} \sqrt{-h} \times
\mathcal{O}(\rho^{-(2d-4)}) = \frac{1}{16\pi G} \int_{\partial M} \mathcal{O}%
(\rho^{-(d-3)}),
\end{equation}
which vanishes for $d \ge 4$. Having shown that $S_{renorm}$
provides a valid variational principle for i) cylindrical temporal
cut-off (${\cal T}^{cyl}$) and either cylindrical or hyperbolic
spatial cut-off ($\Omega^{cyl}$ or $\Omega^{hyp}$) and ii)
hyperbolic temporal cut-off ${\cal T}^{hyp}$) with hyperbolic
spatial cut-off $\Omega^{hyp}$, it is now straightforward to show
that for such cases the numerical value of the action on a
solution is invariant under a change of spatial cut-off of the
form
\begin{equation}
\label{shiftOmega} \Omega \rightarrow \Omega + \delta \Omega,
\end{equation}
with $\delta \Omega = {\cal O}(\rho^0)$; i.e., the value of
$S_{renorm}$ depends at most on the choice of $\Omega^{cyl}$ or
$\Omega^{hyp}$, but not on the precise choice of $\Omega$ within
either class.  To see this, simply note that the change $\delta
h_{ij}$ induced by (\ref{shiftOmega}) takes the form $\delta
h^1_{ij} = 2 \delta \Omega \  h^0_{ij}$, together with changes in
the higher order terms in $h_{ij}$.  Since we have just shown that
$\delta S_{renorm}$ vanishes under any such variation about a
solution, it is clear that that the numerical value of
$S_{renorm}$ is invariant under shifts (\ref{shiftOmega}).

\subsection{Other proposed Counter-terms}

\label{other} Our counter-term $S_{new \ CT}$ is not the first counter-term to have been proposed for asymptotically flat spacetimes. In particular, Mann showed \cite%
{Mann} that for $d=4$ a counter-term proportional to $\sqrt{\mathcal{R}}$
leads to a finite on-shell action for Schwarzschild spacetimes with $\Omega
= \Omega^{cyl}$ and could, in this context, be related to the known
counter-terms \cite{skenderis,kraus} for asymptotically AdS spacetimes. This
result was generalized to arbitrary spacetime dimension by Kraus, Larsen,
and Siebelink \cite{KLS}, who also established similar results for the
counter-term
\begin{equation}  \label{KLS}
S_{KLS} = \frac{1}{8\pi G} \int_{\partial M} \sqrt{-h} \frac{\mathcal{R}%
^{3/2}}{\sqrt{\mathcal{R}^2 - \mathcal{R}_{ij} \mathcal{R}^{ij} } },
\end{equation}
for both cylindrical and hyperbolic boundaries, and in fact for any
boundary metric which agrees to leading order with the standard metric on $%
S^n \times {\mathbb R}^{d-n-1}$. Here we proceed further, considering arbitrary
asymptotically flat vacuum solutions and addressing both the value
of the action and the issue of whether the first variations
vanish.  We consider both (\ref{KLS}) and the counter-term
\begin{equation}  \label{sqrtR}
S_{\sqrt{\mathcal{R}} } = \frac{1}{8\pi G} \sqrt{\frac{n}{n-1}}
\int_{\partial M} \sqrt{-h} \sqrt{\mathcal{R}},
\end{equation}
which generalizes Mann's counter-term (see also \cite{KLS}) to
$S^n \times {\mathbb R}^{d-n-1}$.  Here $S_{\sqrt{\mathcal{R}}}$
depends explicitly on the choice of cut-off $\Omega$ through the
integer $n$.
Due to this feature, it is not clear how $S_{\sqrt {\mathcal{R}}}$ might be usefully generalized away from the above classes of
spatial cut-offs $\Omega$ (e.g., to cut-offs $\Omega = \omega \Omega$ for
smooth non-vanishing functions $\omega$ on the unit hyperboloid).

Let us first consider the counter-terms $S_{\sqrt{\mathcal{R}}}$ and $S_{KLS}$ for cylindrical temporal cut-off ${\cal T}^{cyl}$.  Here each counter-term is only linearly divergent, so in discussing finiteness of the action it suffices to consider the leading order term.  One may readily check that, to leading order, $\hat K$ defined by (\ref{GCM}) agrees with both $%
\sqrt{\frac{n}{n-1} \mathcal{R} }$ and $\frac{\mathcal{R}^{3/2}}{\sqrt{%
\mathcal{R}^2 - \mathcal{R}_{ij} \mathcal{R}^{ij} } }$ for both
cylindrical spatial cut-offs $\Omega^{cyl}$ and hyperbolic spatial
cut-offs $\Omega^{hyp}$. Thus, under these conditions the
counter-terms $S_{\sqrt{\mathcal{R}}}$ and $S_{KLS}$ lead to
renormalized actions that are finite on-shell.

The fact that the counter-terms agree to leading order in $\rho$
is essentially true by construction, as both
$S_{\sqrt{\mathcal{R}}}$ and $S_{KLS}$ were motivated by the fact that they
cancel the leading divergences in the Gibbons-Hawking term. Somewhat
surprisingly, we also find that the first order variations of both $S_{\sqrt{%
\mathcal{R}}}$ and $S_{KLS}$ about such backgrounds exactly match
those of our original $S_{renorm}$ (\ref{Sct}).   The key steps in such calculations
for $S_{\sqrt{\cal R}}$ are:
\begin{eqnarray}
\label{varSR}
\delta \sqrt{\frac{n}{n-1} \mathcal{R} } &=& \frac{1}{2 \sqrt{\mathcal{R}}}
\sqrt{\frac{n}{n-1}} \left( \mathcal{R}_{ij} \delta h^{ij} + \delta \mathcal{%
R}_{ij} h^{ij} \right) \cr &=& - \frac{1}{2}
( \hat K^{ij} + \mathcal{O} (\tilde r^{-d}) )\delta h_{ij},
\end{eqnarray}
where in the last step we have again used (\ref{dR}) to show that, to leading order in $\rho$, the $%
\delta \mathcal{R}_{ij}$ term is a total divergence. We have also used the
Gauss-Codazzi equations (\ref{GCM}) to show that $\mathcal{R}_{ij} = \frac{%
n-1}{n} \hat K \hat K_{ij} + \mathcal{O}(\tilde r^{-(d-3)})$,
where $\tilde r $ is $\rho$ for the hyperboloid, $r$ for the
cylinder ${\mathbb R} \times S^{d-2}$, and the analogous radial
coordinate in the more general case. The analogous calculation for
$S_{KLS}$ yields
\begin{equation}
\label{varKLS}
\delta \frac{\mathcal{R}^{3/2}}{\sqrt{\mathcal{R}^2 - \mathcal{R}_{ij}
\mathcal{R}^{ij} } } = - \frac{1}{2} ( \hat K^{ij} + \mathcal{O}(\tilde
r^{-d})) \delta h_{ij}.
\end{equation}
Comparing with (\ref{v1}) and noting that $\delta \hat K = \delta
\hat K_{ij} h^{ij} + \hat K_{ij} \delta h^{ij}$, we see that for
cylindrical temporal cut-off ${\cal T}^{cyl}$ the counter-terms
$S_{\sqrt{\cal R}}$ and $S_{KLS}$ have the same variations about
solutions as does our new counter-term. As a result, with
cylindrical temporal cut-off ${\cal T}^{cyl}$, both of these
counter-terms again
yield actions which are stationary on solutions, so long as the spatial cut-off induces a boundary of the form $%
S^n \times {\mathbb R}^{d-n-1}$.

Let us now consider hyperbolic temporal cut-offs (${\cal T}^{hyp}$).  In this case each counter-term is quadratically divergent.  Now, we have already established that the counter-terms $S_{\sqrt{\mathcal{R}}}$ and $S_{KLS}$ agree with $S_{new \ CT}$ to leading order in $\rho$.  Furthermore, the behavior at next order in $\rho$ may be considered to be the result of perturbing an original induced metric $h_{ij} =  \rho^2 (h^0)_{ij}$ by $\rho h^1_{ij}$.  Thus, we may compute the next order term using the formulas (\ref{varSR}) and (\ref{varKLS}), which show that they agree with the corresponding expression for $S_{new \ CT}$ up to total derivative terms.  This shows that the counter-terms $S_{\sqrt{\mathcal{R}}}$ and $S_{KLS}$ define a finite action $S_{renorm}$ for hyperbolic temporal cut-off ${\cal T}^{hyp}$ when the spatial cut-off is also chosen to be hyperbolic ($\Omega^{hyp})$.

For $d \ge 5$  and hyperbolic spatial cut-off $\Omega^{hyp}$,
stationarity of $S_{renorm}$ again follows from (\ref{varSR}) and
(\ref{varKLS}). However, as with $S_{new \ CT}$, the situation is
more subtle for $d=4$.  In this case one must calculate the ${\cal
O}(\tilde r^{-d})$ corrections to (\ref{varSR}) and
(\ref{varKLS}).  Such calculations for $S_{\sqrt{\cal R}}$ show
that the corresponding $S_{renorm}$ is {\it not} in general
stationary (see appendix \ref{C}). We have not performed the
corresponding calculations for $S_{KLS}$.  Finally, we note
that in any case where it provides a valid variational principle,
$S_{renorm}$  as defined by $S_{\sqrt{\cal R}}$ or $S_{KLS}$ is
invariant under changes of spatial cut-off of the form
(\ref{shiftOmega}) with $\delta \Omega = {\cal O}(\rho^0)$ by the
same argument used for $S_{new \ CT}$ in section \ref{varSrenorm}.
Namely, (\ref{shiftOmega}) induces a change in $h_{ij}$ equivalent
to one for which we have just shown that $\delta S_{renorm}=0$.

\section{Conserved Quantities}

\label{CQ}

Having constructed a variational principle of the desired type, one may expect that
conserved quantities (e.g., energy, angular momenta) now follow by a
straightforward construction as in Noether's theorem. Such a result is known
\cite{HIM2} in a general setting appropriate to gravitational actions
constructed from a counter-term prescription and modeled on the case of
asymptotically anti-de Sitter spaces. The discussion of \cite{HIM2} is
phrased in terms of ``boundary fields'' and a ``boundary metric,'' but in
the current context we may take our boundary metric to be $\gamma_{ij} =
\lim_{\Omega \rightarrow \infty} \Omega^{-2} h_{ij}$ and all other boundary
fields to be zero. For either spatial cut-off ($\Omega^{hyp}$ or $\Omega^{cyl}$%
), we may choose coordinates on $\partial \mathcal{M}$ such that $%
\gamma_{ij} $ is non-degenerate.

The treatment in \cite{HIM2} is quite general, and does not
specify in detail the way in which either the boundary
manifold $\partial \mathcal{M}$ or the boundary fields are to be
associated with $\mathcal{M}$ and the dynamical fields. In
particular, in our context the $\Omega \rightarrow \infty $ limit
of $\partial \mathcal{M}$ need not be smoothly attached to any
conformal compactification of $\mathcal{M}$, or even to a
compactification of the form described by Ashtekar and Romano
\cite{AshtekarRomano}. What is required is simply that the
boundary fields capture the boundary conditions, and that
diffeomorphisms on $\mathcal{M}$ induce diffeomorphisms on
$\partial M$. It
is clear that both are the case here. The main argument (section III of \cite%
{HIM2}) then considers any asymptotic Killing field $\xi $, where
an asymptotic Killing field is defined to be a vector field which
generates diffeomorphisms that preserve the asymptotic conditions.
In particular, the diffeomorphism should preserve the definition (\ref{AFdef}) of asymptotic flatness and,
for $d=4$ with ($\Omega^{hyp}$, ${\cal T}^{hyp}$) infinitesimal, such diffeomorphisms should result in a variation satisfying
(\ref{varh1}).  Asymptotic symmetries must also preserve  any other condition required to define a proper covariant phase space whose
symplectic structure is both finite and conserved; see, e.g.,
\cite{ABR} for the $d=4$ covariant phase space containing
Minkowski space and a discussion of how the corresponding
restrictions remove both supertranslations and logarithmic
supertranslations \cite{PGB,AshLog} from the list of candidate
asymptotic symmetries.

In such cases, \cite{HIM2} considers the operator $\Delta _{f,\xi
}=\pounds _{f\xi }-f\pounds _{\xi }$ on the space of field
histories, where $\pounds _{\eta }$ denotes the Lie derivative
along the vector field $\eta $ and where $f$ is any smooth bounded
function which vanishes in a neighborhood of the past boundary and
takes the value $f=1$ in a neighborhood of the future boundary.
The argument of \cite{HIM2} then shows that the quantity $-\Delta
_{f,\xi }S_{renorm}$ generates asymptotic diffeomorphisms along
$\xi $ via the Peierls
bracket\footnote{%
The Peierls bracket \cite{Peierls} is a covariant Poisson structure on the
space of solutions modulo gauge transformations. It agrees with the
push-forward of the Poisson bracket of Dirac observables under any evolution
map which takes gauge orbits on the constraint surface in phase space to
such equivalence classes of solutions. See also \cite{Bryce1,Bryce2,Bryce3}
and see \cite{gen} for extensions of the Peierls bracket to algebras of
gauge-dependent quantities and \cite{Fred1,Fred2} for recent related work in
quantum field theory.}.

While we will not repeat the full proof here, we will pause to demonstrate two important
properties of $\Delta _{f,\xi }S_{renorm}$; namely, that it is both finite and differentiable on the space of field histories.   To see these properties,
note first that $\pounds_{f\xi}$ generates a diffeomorphism.  But the space of histories on which $S_{renorm}$ is finite and differentiable is covariant under such diffeomorphisms, so this part of $\Delta_{f,\xi}$ creates no difficulties.   Furthermore,
since $\xi$ is an asymptotic symmetry,
the transformed metric $g_{transformed } =  (1 + \pounds_\xi) g$ satisfies the same
asymptotic conditions as $g$ and, since these conditions are local, so does
$g_{transformed, f } =  (1 + f\pounds_\xi) g$.  Thus,
$(1 + \epsilon \Delta_{f,\xi}) S_{renorm}$ is finite and differentiable.  Finally, since the Lagrangian
is a differentiable function of the fields and their derivatives, $(1 + \epsilon \Delta_{f,\xi})S_{renorm}$ will
also be linear in $\epsilon$.  Thus
$\Delta_{f,\xi} S_{renorm}$ is  well-defined, finite, and  differentiable on the chosen space
of histories.

Now, since $-\Delta _{f,\xi
}S_{renorm}$  generates asymptotic $\xi$-translations, it
can differ from any Hamiltonian definition of the conserved
quantity associated to $\xi$ by at most a ``c-number;'' i.e., by a quantity having trivial Peierls bracket with any observable.  Such a quantity may be a function of
the auxiliary structure $\Omega $, but must be otherwise constant over
the space of solutions. In particular, since the usual ADM charges $H_{ADM}[\xi]$
\cite{ADM1,ADM2,ADM3} vanish on Minkowski space,  given any metric $g$ on $\mathcal{M}$%
, if we define
\begin{equation}
\label{Q}
Q[\xi] =  - \Delta_{f,\xi}S_{renorm}[g] \ \ \ {\rm and} \ \ \ Q_0[\xi] = - \Delta_{f,\xi}S_{renorm}[g_{Mink}],
\end{equation}
where $g_{Mink}$ is the Minkowski metric, then $Q[\xi]-Q_0[\xi]$ must agree with the corresponding $H_{ADM}[\xi]$ (or, in fact, with any other standard definition such as \cite%
{WaldBook,AshtekarHansen,AshtekarInf,AshtekarRomano,RT,AD,SD2,SD3} of the charge)
 whenever these more familiar charges are well-defined.
We note, however, that (\ref{Q}) also extends such definitions to
allow for a larger magnetic Weyl tensor at infinity and, in
particular, to the case of non-vanishing NUT charge in 4 spacetime
dimensions.

Now, the full discussion of \cite{HIM2} does use some structure that is
\textit{not} present in our case in order to derive additional results.
These additional results would relate $-\Delta_{f,\xi}S_{renorm}$ to a
`boundary stress tensor' given by variations of the
action $S_{renorm}$ with respect to
$\gamma_{ij}$. In the current context, such variations turn out to diverge.
However, working for the moment with the regulated action $S_{renorm, \Omega}$ associated with a cut-off region of spacetime ${\cal M}_\Omega$ for some finite value of $\Omega$%
, we may follow \cite{BY} and define a `boundary stress tensor' as a function\footnote{%
For a given spacetime, this function will be defined for values of $\Omega$ large enough that  equation (\ref{GCM})
can be solved for $\hat K_{ij}$ so that our counter-term is well-defined. Recall that
our counter-term need not be well-defined
for small $\Omega$.  } of $\Omega$ through variations of $S_{renorm,\Omega}$
with respect to $(h_\Omega)_{ij}$. This stress tensor admits an expansion of
the form
\begin{eqnarray}  \label{Texp}
T_{ij} (\Omega)&:=& \frac{-2}{\sqrt{-h}} \frac{\delta S_{renorm, \Omega} }{%
\delta h_\Omega^{ij}} = \frac{1}{16 \pi G} (\pi_{ij} - \hat{\pi}_{ij}) \cr %
&:=& \Omega^{-(d-4)} \left( T_{ij}^{0} + \Omega^{-1} T_{ij}^1 + \ terms \
vanishing \ faster \ than \ \Omega^{-1} \right).
\end{eqnarray}

Note that, since it is local on $\partial \mathcal{M}_\Omega$, the
definition of $T_{ij}(\Omega)$ depends only on the spatial cut-off
and is independent of the choice off temporal cut-off
($\mathcal{T}^{cyl}$ or $\mathcal{T}^{hyp}$). In general, the
definition of $T_{ij}$ will depend on the precise choice of
counter-term ($S_{new \ CT}$, $S_{sqrt{\cal R}}$, or $S_{KLS}$) and the choice of
spatial cut-off.
For definiteness, we will fix our attention on the counter-term
$S_{new \ CT}$  below. In this case, for the
hyperbolic spatial cut-off ($\Omega^{hyp}$) we see from (\ref{step3}) that
$T^0_{ij}$ is proportional to the leading term in the electric
part of the Weyl tensor:
\begin{equation}
\label{TE}
T^0_{ij} = \lim_{\Omega \rightarrow \infty}\frac{1}{8\pi G} \frac{\rho^{d-3}%
}{d-3} E_{ij} ;
\end{equation}
Derivation and discussion of the corresponding expression for $T^1_{ij}$ will be left for future study \cite{MMnext}.

Despite working at finite $\Omega$, conservation of this boundary stress tensor follows from the usual argument: One notes that the cut-off
action $S_{renorm, \Omega}$ is invariant under diffeomorphisms of $\mathcal{M%
}_\Omega$ preserving $\partial \mathcal{M}_\Omega$. Since any
diffeomorphism of $\partial \mathcal{M}_\Omega$ can be extended
to such a diffeomorphism of $\mathcal{M}_\Omega$, it follows that
$\delta S_{renorm, \Omega} =0$ whenever $(\delta h_\Omega)_{ij} =
D_{(i} \xi_{j)}$ for any vector field $\xi^j$ on $\partial
\mathcal{M}$. Thus, taking a variational derivative of $S$ with
respect to such $\xi^j$ shows that $T_{ij} (\Omega)$ is conserved
at each $\Omega$; i.e., that
\begin{equation}
\label{Tcons}
D^i T_{ij} (\Omega) = 0.
\end{equation}
In particular, equation (\ref{Tcons}) holds separately for each finite value of $\Omega$.
It turns out that the same general steps as in \cite{HIM2} allow us to relate $Q[\xi]$ to the boundary stress tensor.  The argument below holds for either class of spatial cut-off ($\Omega^{hyp}$ or $\Omega^{cyl}$), though we remind the reader that our coordinates $\eta^i$ are defined in terms
of the hyperboloid.  It also holds for either class of temporal cut-offs (${\cal T}^{cyl}$ or ${\cal T}^{hyp}$).

Let us begin by computing $Q[\xi]$ in terms of variations of the action:
\begin{equation}
\label{Q2}
Q[\xi] := - \Delta_{f,\xi} S_{renorm} = - \lim_{\Omega \rightarrow \infty} \left( \int_{{\cal M}_\Omega} \frac{\delta S_{renorm}}{\delta g_{ab}}
\Delta_{f,\xi} g_{ab} + \frac{1}{2} \int_{\partial {\cal M}_\Omega} \sqrt{-h_\Omega} T^{ij} \Delta_{f,\xi} h_{ij}.
\right)
\end{equation}
Note that the bulk term vanishes, as we evaluate $Q[\xi]$ on a solution.  Furthermore, it is straightforward
to calculate $\Delta_{f,\xi} h_{ij}$:
\begin{equation}
\label{evalDh}
\Delta_{f,\xi} h_{ij} =  (\pounds_{f\xi} g)_{ij} - f  (\pounds_{\xi} g)_{ij} =  \xi_i D_j f +  \xi_j D_i f,
\end{equation}
where $(\pounds_{f\xi} g)_{ij}$ and $(\pounds_{\xi} g)_{ij}$
denote quantities evaluated in the bulk of ${\cal M}_\Omega$ and
then pulled-back to the boundary $\partial {\cal M}_\Omega$. We
may now use (\ref{evalDh}) to write (\ref{Q2}) in the form
\begin{equation}
\label{Q3}
Q[\xi] = - \lim_{\Omega \rightarrow \infty}  \int_{\partial {\cal M}_\Omega}   \Omega^{-(d-3)}  \sqrt{-h} \left(\Omega  T^0_{ij}
 +  T^1_{ij} \right) h^{jk} h^{il}
 \xi_l D_k f,
 \end{equation}
 where we have dropped the higher terms in the expansion of the boundary stress tensor;
 since $\xi_i$ is largest at infinity for a boost, which has $\xi^i = {\cal O} (\rho^0)$ and
 $\xi_i = {\cal O} (\rho^2)$, such higher terms in
 $T^{ij}(\Omega)$
 will not contribute to $Q[\xi]$ in the limit $\Omega \rightarrow \infty$ for either class of temporal cut-offs
 ${\cal T}^{hyp}$ or ${\cal T}^{cyl}$.

 In fact, the overall scaling of terms in (\ref{Q3}) is the same for both ${\cal T}^{hyp}$ and ${\cal T}^{cyl}$: while for
${\cal T}^{cyl}$, the coordinate volume of $\partial {\cal M}_\Omega$ scales as $1/\Omega$, but the
 time derivative of $f$ scales with a compensating factor of
$\Omega.$
 Integrating (\ref{Q3}) by parts, we may write our charge as\footnote{The rest of this section makes use of the global structure of (\ref{AFdef}), and so does not directly apply to spacetimes with Lorentzian NUT charge.  In particular, the boundary of a NUT-charged spacetime does not admit a global cross-section ${\cal C}$.
However, instead of integrating over a cross-section, one can use the fact that smooth spacetimes with Lorentzian NUT charge are periodic in time to write expressions similar to those below involving integration over all of $\partial {\cal M}$.}
\begin{equation}
\label{Q4} Q[\xi] =
 \lim_{\Omega \rightarrow \infty}
 \int_{C_{\Omega }}  \Omega^{-(d-3)} \sqrt{h_{C_{\Omega}}}  \left(\Omega  T^0_{ij}
 +  T^1_{ij} \right)  h^{il}
\xi_l n_\Omega^j + Q_{vol}[\xi],
\end{equation}
where $C_\Omega = \Sigma_+ \cap \partial{\cal M}_\Omega$ is the
future boundary of
 $\partial{\cal M}_\Omega$, $n^j_\Omega$ is its future (i.e., outward)-pointing unit normal
 in $\partial {\cal M}_\Omega$, and $h_{C_\Omega}$ is the induced metric on $C_\Omega$.
In addition, we have separated off the `volume term' which
includes an integral over all of $\partial {\cal M}_\Omega$:
\begin{equation}
 Q_{vol}[\xi] :=
 \lim_{\Omega \rightarrow \infty}
 \int_{\partial {\cal M}_\Omega} f \Omega^{-(d-3)}  \sqrt{-h} \left(\Omega  T^0_{ij}
 +  T^1_{ij} \right) h^{jk} h^{il}
 D_k \xi_l .
 \end{equation}

The first term in (\ref{Q4}) is of the sort used to define charges
in the asymptotically anti-de Sitter context via the so-called
boundary counter-term method \cite{skenderis,kraus}.  This term
could itself be taken to give a (slightly different) definition of
the charge. Indeed, the second term $Q_{vol}[\xi]$ can be shown to
be a ``c-number,'' meaning that it Peierls-commutes with all
observables.  Thus, the first term in (\ref{Q4}) alone also
generates asymptotic $\xi$-translations. To see that
$Q_{vol}[\xi]$ is a c-number, recall that $Q[\xi]$ generates the
asymptotic symmetry $\xi$ for any function $f$ which vanishes on
the past boundary and satisfies $f=1$ on the future boundary.
Thus, the difference $Q[\xi; f_1]-Q[\xi;f_2]$ must be a c-number whenever $Q[\xi; f_1]$ and $Q[\xi;
f_2]$ are charges of the form (\ref{Q})  defined by two such functions $f_1, f_2$.  In computing
such a difference from (\ref{Q4}), the first term cancels and we
are left only with the difference $Q_{vol}[\xi; f_1] - Q_{vol}[\xi; f_2]$.
But since $D_k \xi_l + D_l \xi_k = {\cal O}(\rho^{-(d-5)})$, it is
clear that $Q_{vol}[\xi;f_2]$ vanishes in the limit where $f_2$ is zero
everywhere except within a tiny neighborhood of the future
boundary. Thus, for any allowed $f_1$, we see that  $Q_{vol}[\xi; f_1]$ must be
a c-number.

As a result, $Q_{vol}[\xi]$ can in general depend at most on the auxiliary
structure and the particular choice of covariant phase space
(i.e., on the choice of boundary conditions) and must be constant
over any given covariant phase space. Suppose then that we choose
a covariant phase space which contains Minkowski
space. Within this context, the second term in (\ref{Q4}) is given
by its value on Minkowski space itself.  But the Gauss-Codazzi
equations guarantee that $\hat K_{ij} = K_{ij}$, and thus that
both $T_{ij}(\Omega)$ and (\ref{Q4}) vanish identically (for
either $\Omega^{hyp}$ or $\Omega^{cyl}$).    We therefore see that
our charges $Q[\xi]$ agree precisely with the usual definitions
\cite{AshtekarHansen,ABR,AshtekarRomano,RT,ADM1,ADM2,ADM3} on this
covariant phase space.

In a more general context, $Q_{vol}[\xi]$  will
still vanish for hyperbolic spatial cut-off $\Omega^{hyp}$.
Because this cut-off is invariant under boosts, we see that $D_k
\xi_l + D_l \xi_k =  \frac{\alpha}{\rho} h^0_{ij}$ for some smooth
function $\alpha$ on ${\cal H}^{d-1}$.  Since for $\Omega^{hyp}$
the leading stress-tensor term $T^0_{ij}$ is traceless with
respect to $h^0_{ij}$, this is sufficient to make $Q_{vol}[\xi]$
 vanish.  Similarly, for cylindrical
spatial cut-off $\Omega^{cyl}$, $Q_{vol}[\xi]$ must vanish unless
$\xi^i$ contains a boost.  The difficulty with boosts is that
while, in analogy with $\Omega^{hyp}$, we have $D_k \xi_l + D_l \xi_k = \beta{\xi}h^0_{ij}$ for some smooth
function $\beta$ on ${\cal H}^{d-1}$,
for $\Omega^{cyl}$ the leading stress-tensor term $T^0_{ij}$ fails to be
traceless.

When $Q_{vol}[\xi]$ vanishes, we may formally write (\ref{Q4}) as
\begin{equation}
\label{Qhyp}
Q[\xi] =
 \int_{C}  \sqrt{h_{C}}  T_{ij} \xi^i n^j,
\end{equation}
where $C = \Sigma_- \cap \partial {\cal M}$ and
$(h_C)_{ij}, n^j$ are the associated induced metric and
future-pointing normal vector field.  In fact, the vanishing of
$Q_{vol}[\xi]$ for all $f$ is sufficient to guarantee that the
charge (\ref{Qhyp}) conserved in the sense that its numerical
value is independent of the cut $C$. Thus, for the particular
cases described above, we have much of the structure which has
become familiar \cite{skenderis,kraus} in the anti-de Sitter
context and, in particular, the conserved charges may be
calculated by an algorithm of the sort described in
\cite{Mann2,DAR}.

For hyperbolic spatial cut-off $\Omega^{hyp}$, using the relation
(\ref{TE}) between $T^0_{ij}$ and the electric part of the Weyl
tensor, expression (\ref{Qhyp}) makes manifest the general
agreement with standard ADM expressions for energy and momentum\footnote{In addition, comparing (\ref{Q4}), (\ref{Qhyp}) with the results of  \cite{AshtekarHansen,AshtekarInf, AshtekarRomano} for angular momentum strongly suggests, at least
for $d=4$, for spatial cut-off $\Omega^{hyp}$, and when the asymptotic metric takes the form specified in
\cite{AshtekarHansen,AshtekarInf, AshtekarRomano}, that $T^1_{ij}$ may be expressed in terms of the magnetic part of the Weyl tensor.},
as the latter are known from \cite{AshtekarHansen,AshtekarInf,
AshtekarRomano} to be expressible via a relation analogous to
(\ref{Qhyp}) in terms of the leading contribution to $E_{ij}$.
Now, counting powers of $\Omega$ in (\ref{Qhyp}) may cause the
reader to worry that a divergent contribution arises from the term
involving $\hat \xi^i_B$ and $T_0^{ij}$.  However, we remind the
reader that we have already shown $Q[\xi]$ to be finite.  In $d=4$
spacetime dimensions, one may show
\cite{AshtekarHansen,AshtekarInf, AshtekarRomano} that the
integral over $C$ of the apparently divergent term vanishes due to
the fact that the leading contribution to $E_{ij}$ admits a
certain scalar potential.  In higher dimensions, it is clear that
some analogous cancellation must occur.

Note that the above argument also shows that (\ref{Qhyp}) holds
for the stress tensors computed from $S_{\sqrt{\cal R}}$ and from
$S_{KLS}$.  Furthermore, from the definition (\ref{GCM}) of $\hat
K_{ij}$, it is straightforward to show that the stress tensor
defined by $S_{new \ CT}$, $S_{\sqrt{\cal R}}$, and $S_{KLS}$
differ only by terms built from $h^1_{ij}$ and higher order terms
in $h_{ij}$.  Thus, these stress tensors all agree (and all
vanish) for Minkowski space.  Thus, we see that the conserved
quantities $Q[\xi]$ as defined through $S_{\sqrt{\cal R}}$ or
$S_{KLS}$ also agree with the standard results
\cite{AshtekarHansen,ABR,AshtekarRomano,RT,ADM1,ADM2,ADM3}. In
particular, this observation justifies the calculations performed
in \cite{DAR}.

\section{Discussion}

\label{disc} In the above work, we considered actions $S_{renorm}$
for asymptotically flat vacuum gravity, constructed by adding one
of the counter-terms $S_{new \ CT}$ (\ref{Sct}),
$S_{\sqrt{\mathcal{R}}}$ (\ref{sqrtR}) \cite{Mann} , or $S_{KLS}$
(\ref{KLS}) \cite{KLS} to the Einstein-Hilbert action with
Gibbons-Hawking term. All of these counter-terms are given by
local algebraic functions of the boundary metric and Ricci tensor.
The new counter-term $S_{new \ CT}$ is constructed from a
symmetric tensor $\hat K_{ij}$ defined by solving the
traced Gauss-Codazzi equations (\ref{GCM}) which would result if the boundary spacetime ($%
\partial \mathcal{M}, h$) were detached from the bulk spacetime $(\mathcal{M}%
,g)$ and isometrically embedded in Minkowski space. As a result,
$S_{new \ CT}$ is directly related to the reference background
counter-term $S_{Ref}$, though with the advantage that, as opposed
to $S_{Ref}$, the new counter-term $S_{new \ CT}$ is well-defined
on open sets of boundary metrics in dimesions $d \ge 4$.  

The results established depend on the choice of temporal cut-off used to define the system.
The simplest case is that of a cylindrical temporal cut-off (${\cal T}^{cyl}$).  For this case, we have demonstrated that each of the three counter-terms $S_{new \ CT}, S_{%
\sqrt{\mathcal{R}}}, S_{KLS}$ leads to an action $S_{renorm}$ such that

\begin{itemize}

\item  For $d \ge 4$ the action $S_{renorm}$ leads to a fully satisfactory variational principle when $\partial {\cal M}$ is defined by
either a hyperbolic or a cylindrical spatial cut-off; i.e., by  $\Omega^{hyp}$ (\ref{hypcut})  or $\Omega^{cyl}$  (\ref{cylcut}).  In particular, $S_{renorm}$ is both finite  and stationary on asymptotically flat vacuum solutions. In addition, on any such solution the action takes a numerical value which is  independent of the precise choice of spatial cut-off within the class $\Omega^{hyp}$
(\ref{hypcut})  or $\Omega^{cyl}$  (\ref{cylcut}).
\end{itemize}

Also,
\begin{itemize}
\item For $d \ge 4$, the action $S_{renorm}$ defined by the
counter-term $S_{new \ CT}$ is again finite when $\partial {\cal
M}$ is defined by any spatial cut-off of the form $\Omega = \omega
\Omega^{hyp}$ for $\omega$ a smooth non-vanishing function on the
unit hyperboloid. Again, the numerical value of the action depends
at most on the choice of this $\omega$ and not on further details
of the cut-off. However, except as stated above, we reached no
conclusions with regard to the variations of $S_{renorm}$ in this
context.  We note that the counter-term $S_{\sqrt{\cal R}}$ is not
defined for such general spatial cut-offs, and we have not
investigated the corresponding result for $S_{KLS}$.
\end{itemize}

When a  hyperbolic temporal cut-off (${\cal T}^{hyp}$) is chosen, both the Gibbons-Hawking term
and the counter-terms are more divergent.  Thus, the analysis is more subtle.  In this case, we have established that

\begin{itemize}
\item  When defined using our new counter-term $S_{new \ CT}$,
the action $S_{renorm}$ leads to a fully satisfactory variational principle for $d \ge 4$ and when $\partial {\cal M}$ is defined by a hyperbolic spatial cut-off; i.e., by  $\Omega^{hyp}$ (\ref{hypcut}).  In particular, $S_{renorm}$ is both finite  and stationary on asymptotically flat vacuum solutions. In addition, on any such solution the action takes a numerrical value which is  independent of the precise choice of spatial cut-off within the class $\Omega^{hyp}$
(\ref{hypcut}).

\item For the remaining counter-terms $S_{%
\sqrt{\mathcal{R}}}$ and $S_{KLS}$, the action $S_{renorm}$ is
again finite on vacuum solutions for $d \ge 4$ when when $\partial
{\cal M}$ is defined by a hyperbolic spatial cut-off; i.e., by
$\Omega^{hyp}$ (\ref{hypcut}). However, we have demonstrated that
the action is stationary only for $d \ge 5$.  For $S_{\sqrt{\cal
R}}$, we have demonstrated that the action is {\it not} in general
stationary for $d =4$, while we have not performed the calculation
for $S_{KLS}$ in $d=4$ to the relevant order in $\rho$. 
However, whenever these actions are stationary on such solutions,
the numerical value of $S_{renorm}$ is again independent of the
particular spatial cut-off $\Omega$ chosen within the class
$\Omega^{hyp}$ (\ref{hypcut}).

\end{itemize}

All of these results hold up to possible terms localized at the past and future boundaries ($\Sigma_-$ and $\Sigma_+$), as such terms have been neglected in our treatment.  It is clearly of interest to address such terms in future work.

While our focus has been on Lorentz signature, analogous results follow immediately in the Euclidean
setting.  We note that, in the thermal context, periodicity of Euclidean time automatically imposes a  cylindrical temporal cut-off and removes the possibility of boundary terms on $\Sigma_+$ and $\Sigma_-$.

 Under the conditions stated above for which each action
$S_{renorm}$ is both finite and stationary, we showed that
$S_{renorm}$ leads via a straightforward algorithm to the usual
conserved quantities
 $Q[\xi]$ (energy, angular momentum, etc.) at spatial infinity for each asymptotic symmetry $\xi$.
 We find this to be a
significant conceptual simplification over the textbook
calculations of such quantities. Furthermore, our construction
generalizes the usual definitions
to spacetimes in which the magnetic and electric parts of
the Weyl tensor are of the same order and, in particular, for
non-zero NUT charge\footnote{A canonical definition of conserved
charges in the presence of NUT charge will appear soon
\cite{HTnew}.  Due to the Peierls argument of \cite{HIM2}, the
canonical definitions should agree with the covariant ones given
here.  However, we remind the reader that we have not actually
constructed the phase space for such solutions here, and have thus
not kept track of any conditions required to ensure that the
symplectic structure is finite.  As a result, though given any
asymptotic symmetry our results will yield the correct conserved
quantity, we have not determined the precise asymptotic symmetries
of any system. Instead, these are taken as an input in the present
work.} when $d=4$.  We have also shown that our quantities can be
expressed in terms of a `boundary stress tensor,' though both the
leading term ($T^0_{ij}$) and a sub-leading term ($T^1_{ij}$) are
required to construct all conserved quantities. In most cases,
$Q[\xi]$ can be expressed in the form
\begin{equation}
Q[\xi] = -
 \int_{C}  \sqrt{h_{C}}  T_{ij} \xi^i n^j,
\end{equation}
with the one (possible) exception occurring when $\xi$ generates a
non-trivial boost in the presence of cylindrical spatial cut-off
$\Omega^{cyl}$ with boundary conditions chosen such that Minkowski
space is {\it not} part of the covariant phase space. Thus we have
much of the structure which has become familiar
\cite{skenderis,kraus} in the anti-de Sitter context, and which is
of much use in the AdS/CFT correspondence (see e.g.,
\cite{Juan,MAGOO}).

One notable difference from the anti-de Sitter case is, however,
that we have established the above properties only when the
leading behavior of the metric at infinity has a specific form
(\ref{AFdef}), where in particular $h^0_{ij}$ is an $SO(d-1,1)$
invariant metric on the unit hyperboloid ${\cal H}^{d-1}$. Though
we have not discussed it here, a major obstacle to considering
other $h^0_{ij}$ arises from the form of the Einstein equations
themselves near $i^0$.  In particular, suppose for the moment that
we attempt to allow $h^0_{ij}$ in to be an arbitrary Lorentz
signature boundary metric on ${\mathbb R} \times S^{d-2}$ but to otherwise leave our definition (\ref{AFdef})  of asymptotic flatness unchanged.  This might seem natural if one sought a non-gravitating theory associated with $i^0$ and dual to asymptotically flat gravity\footnote{In contrast, see \cite{AFD1}-\cite{AFD5} for work considering a possible dual theory at null infinity.}. We will, of course, wish to impose the
Einstein equations.  However, it turns out that the Einstein equations alone require
 $h^0_{ij}$ to be an Einstein
metric \cite{AshtekarRomano,BS,B,skenAF}, which for $d=4$ implies
that it is in fact a constant multiple of the metric on the unit
hyperboloid ${\cal H}^3$. Using any other metric for $h^0_{ij}$
would thus necessitate the inclusion of more singular terms in our
ansatz for the metric, and such terms appear difficult to control.

A less ambitious goal would be to maintain the asymptotic conditions (i.e., (\ref{AFdef})) used in this work, but to allow more general spatial cut-offs $\Omega$.  For example,  we have noted that, for spacetimes (\ref{AFdef}), with the
counter-term $S_{new \ CT}$ and cylindrical temporal cut-off $\mathcal{T}%
^{cyl}$, the action $S_{renorm}$ continues to take finite values on shell
for a general class of spatial cut-offs $\Omega$. However, we have not been
able to establish this result for hyperbolic temporal cut-offs $\mathcal{T}%
^{hyp}$, nor have we established that the action is fully stationary on
asymptotically flat solutions with cylindrical temporal cut-off $\mathcal{T}%
^{cyl}$. It would be interesting to explore this further, and also to
investigate the corresponding properties of
 $S_{KLS}$ (\ref{KLS}).

Finally, we note that our discussion has been restricted to
asymptotically flat spacetimes in dimensions $d \ge 4$. The case
$d=3$ is also of significant interest with `asymptotically flat'
boundary conditions such as those of \cite{AV}, as are various other boundary conditions for $d \ge 4$. A particularly interesting case is that of asymptotically Melvin spacetimes \cite{Melvin,Ernst} associated
with black hole pair creation \cite{Gibbons,GS,DGGH}.
The action for such spacetimes is known to be finite when defined
by either  the reference background subtraction prescription
\cite{HawHor}  or the counter-term $S_{\sqrt{\mathcal{R} }}$
\cite{Radu}.  However, the status of the variational principle has
not been addressed.  It would be very interesting to discover if
$S_{new \ CT}$ can provide a suitable variational principle in
such a context.

\appendix

\section{On the restriction of the of the variations for ${\cal T}^{hyp}$ and $d=4$.}
\label{restrict}

In section \ref{prelim} we imposed the restriction $\delta
h^1_{ij} = \alpha h^0_{ij}$ (\ref{varh1}) on variations about
asymptotically flat solutions in $d=4$ spacetime dimensions.
Here $\alpha$ is some
smooth function on the hyperboloid.  The
purpose of this appendix is to argue that this restriction is
compatible with the equations of motion.  By this we mean that, if
an infinitesimal tangent vector to the space of solutions has
initial data satisfying (\ref{varh1}) on some Cauchy surface $C$
of ${\cal H}^{d-1}$, then the tangent vector can be taken to have
this form on all of ${\cal H}^{d-1}$. 

Furthermore, we will show that the initial data of this form is
sufficiently general. We choose our criterion for ``sufficient
generality'' by comparison with \cite{ABR}, which allowed
only one degree of freedom in $h^1_{ij}$ and, furthermore,
imposed a single relation between such $h^1_{ij}$ and $\sigma$.
What we show below is that for any variation satisfying both
$\delta h^1_{ij} = \alpha h^0_{ij}$ and $\delta \sigma =
-\frac{1}{2}\alpha$, all equations of motion hold to the
relevant order in $\rho$ when $\delta \sigma$ satisfies its
equation of motion:
\begin{equation}
\label{sigeq}
D^2 \delta \sigma + 3 \delta \sigma/\rho^2 = \ {\rm higher \ order \ in} \  \rho,
\end{equation}
see equation (3.29) of \cite{BS}.
We base our argument on the results of \cite{BS,B}, who studied
the $d=4$ Einstein equations expanded near spatial infinity.  They
found that each correction $h^n_{ij}$ satisfies an equation on
${\cal H}^{d-1}$ of the form:
\begin{equation}
L_n h^n_{ij} = s_{ij}^n, \end{equation} where $L_n$ is a
hyperbolic linear partial differential operator and $s_{ij}^n$ is
a source term built (not necessarily linearly) from $h^m_{ij}$ for
$m < n$, as well as from $\sigma$ and the corresponding higher
corrections.
  Equation (\ref{sigeq}) is the corresponding equation for
$\sigma$, and there is a similar hierarchy  of equations for
corrections to $\sigma$, as well as certain constraints.
However, it was shown that this system of equations has solutions
whenever the constraints are satisfied on some Cauchy surface of
${\cal H}^{d-1}$. Infinitesimal tangent vectors satisfy the
linearization of this system of equations.

We will now show that, given initial data of the form 
$\delta h^1_{ij} = \alpha h^0_{ij}, \delta \sigma =- \frac{1}{2}
\alpha$ for a tangent vector to the space of solutions, there is
a solution to the (linearization of the) equations of \cite{BS,B},
which again takes the form $\delta h^1_{ij} = \alpha
h^0_{ij}, \delta \sigma= - \frac{1}{2}\alpha$. As we have already
imposed (\ref{sigeq}), the only obstacles are the equation
involving $L_1 h^1_{ij}$ and the corresponding constraints.  We
begin by transcribing the linearization of the equation of motion
involving $L_1 \delta h^1_{ij}$ (see (3.28) of \cite{BS}), as
\begin{eqnarray}
\label{h1eq}
 \frac{1}{2} D_k  D^k \delta h^1_{ij} &-&
\frac{1}{2} D_i  D_j \delta h^1 - 3  \rho^{-2} D_i  D_j
\delta \sigma - 3   \rho^{-2} \delta \sigma \ h^0_{ij} - \frac{3}{2}( \rho^{-2} \delta h^1_{ij} - \frac{1}{3}
\delta h^1 h^0_{ij}) \cr &=& \ {\rm higher \ order \ in} \  \rho,
\end{eqnarray}
where to the desired order $\delta h^1 = (h^0)^{ij} \delta
h^1_{ij}$. Substituting  $\delta h^1_{ij} = \alpha h^0_{ij},
\delta \sigma =-\frac{1}{2} \alpha$ , one finds 
\begin{equation}
\frac{1}{2} h^0_{ij} \left( D_i D^i \alpha + 3 \alpha/\rho^2
\right) =0,
\end{equation}  where in the last step we have used the
linearization of (\ref{sigeq}) with $\delta \sigma = -\frac{1}{2}\alpha$.

It remains only to check the constraint equations.  When
linearized, these become: 
\begin{equation} 
- D^2 \delta h^1+ D^iD^j \delta h^1_{ij}  + 12 \frac{\delta\sigma}{\rho^4} = -\frac{2}{\rho^2} (D^2\alpha + 3
\alpha/\rho^2) = 0,
\end{equation}
 and 
\begin{equation}
- \frac{1}{2} D_i  (\delta h^1)^i_j + \frac{1}{2} D_j \delta h^1
+ 2 D_j \delta \sigma = 0.
\end{equation}
 Thus, we see that, to this order in $\rho$, solutions to the
linearized equations of motion exist with any initial data of the
form  $\delta h^1_{ij} = \alpha h^0_{ij}, \delta \sigma
=-\frac{1}{2} \alpha$.

\section{Sub-leading terms in the variation $\delta S_{new \ CT}$}
\label{ErrorTerm}

This appendix outlines the calculation that
\begin{equation}
\label{answer} \delta \hat K_{ij} h^{ij} = -
\frac{1}{2} \left[ \hat K_{ij} \delta h^{ij} - \frac{\rho}{(d-2)}
h^{ij} \delta {\cal R}_{ij} \right] + {\cal O}(\rho^{-4}),
\end{equation}
for $d=4$ with hyperbolic spatial cut-off $\Omega^{hyp}$ 
and $\delta h^1_{ij} = \alpha h^0_{ij}$. This result establishes
that our new counter-term yields an action $S_{renorm}$ which is
in fact stationary under such conditions. For much of this
calculation it will be convenient to work in general dimension
$d$, though we treat only the case of hyperbolic spatial cut-off
$\Omega^{hyp}$.

Let us begin by defining
\begin{equation}
\label{newexp}
\epsilon_{ij} = \hat K_{ij} - \frac{1}{\rho} h_{ij}.
\end{equation}
Note that $\epsilon_{ij}$ is of order ${\cal O} (\rho^0)$, and that it receives contributions at this order from both  $\hat K_{ij}$ and
$h_{ij}$.  For $\epsilon =0$, equation (\ref{toSolve}) is easy to solve and gives exactly
$\delta \hat K_{ij} h^{ij} = -
\frac{1}{2} ( \hat K_{ij} \delta h^{ij} - \frac{\rho}{(d-2)}
h^{ij} \delta {\cal R}_{ij}) $.  Thus, our task is to solve (\ref{toSolve}) to first order in epsilon.
As with all other objects, indices on $\epsilon_{ij}$ will be raised and lowered with $h_{ij}$.  In particular, $\epsilon : = \epsilon_{ij} h^{ij}$.

To do so, write (\ref{toSolve}) as
\begin{equation}
L_{ij}{}^{kl} \delta \hat K_{kl} = \delta {\cal R}_{ij} - M_{ijkl} \delta h^{kl}.
\end{equation}
We now expand
\begin{eqnarray}
L = L^0 + L^1 + \cdots, \\
L^{-1} = (L^{-1})^0 + (L^{-1})^1 + \cdots, \\
M = M^0 + M^1 + \cdots,
\end{eqnarray}
where terms with superscripts $n$ are homogeneous in $\epsilon_{ij}$ of order $n$ and where
$(L^{-1})_{ij}{}^{kl} (L)_{kl}{}^{mn} = \delta_i^m \delta_j^n.$

One finds
\begin{eqnarray}
[(L^{-1})^0]_{ij}{}^{kl} &=& \frac{\rho}{d-3} \left[ \delta_i^k \delta_j^l - \frac{1}{2(d-2)} h^{kl}h_{ij}\right], \\
h^{ij} [(L^{-1})^1]_{ij}{}^{kl} &=& \frac{\rho^2}{(d-2)(d-3)} \left[  \epsilon^{kl} -  \frac{1}{2} \epsilon h^{kl}\right],\\
M^0_{ijkl} &=& \rho^{-2} (h_{ij} h_{kl} - h_{ik} h_{jl}), \\
h^{ij} M^1_{ijkl} &=& \rho^{-1}\left( \epsilon h_{kl} + (d-3) \epsilon_{kl} \right), \ {\rm or}, \\
h^{ij} M_{ijkl}  &=&  \frac{(d-2)}{\rho} \hat K_{kl} +  \frac{\epsilon h_{kl}}{\rho} 
- \frac{\epsilon_{kl}}{\rho} + {\cal O}(\epsilon^2).
\end{eqnarray}
As a result,
\begin{eqnarray}
\label{xyz}
h^{ij} \delta \hat K_{ij} &=&  \frac{\rho}{2(d-2)} \delta {\cal R}_{ij} h^{ij} - \frac{1}{2} \hat K_{ij} \delta h^{ij} \nonumber \\
&-& \frac{\rho^2}{(d-2)(d-3)} \left[\frac{1}{2} \epsilon \ \delta {\cal R}_{ij} h^{ij} - \epsilon^{ij} \delta {\cal R}_{ij} \right] \nonumber \\
&-& \frac{\epsilon h_{ij} \delta h^{ij}}{2(d-2)(d-3)} + \frac{(d-1) \epsilon_{ij} \delta h^{ij}}{2(d-2)(d-3)} + {\cal O}(\rho^{-(3d-8)}).
\end{eqnarray}

Next we observe that, to the desired order in $\rho$, in any term in (\ref{almostthere}) containing $\epsilon$, we may substitute
\begin{eqnarray}
\label{dhsub}
\delta h_{ij} &\rightarrow& \frac{\delta h^1_{ij}}{\rho^{d-5}} =  \frac{\alpha h^0_{ij}}{\rho^{d-5}} \rightarrow \frac{\alpha h_{ij}}{\rho^{d-3}}, \cr
\delta h^{ij} &\rightarrow& -  \frac{\alpha h^{ij}}{\rho^{d-3}}.
\end{eqnarray}
where we have used (\ref{varh1}).  In addition, we recall that $\delta {\cal R}_{ij}$ is given by expression (\ref{dR}) in terms of second covariant derivatives ($D_k$) of $\delta h_{ij}$.  Thus, in any term containing $\epsilon$ we may substitute
\begin{equation}
\delta {\cal R}_{ij} \rightarrow - \frac{1}{2\rho^{(d-3)}} \left[ (d-3) D_iD_j \alpha + D^2 \alpha \ h_{ij} \right].
\end{equation}

Recalling that we wish to integrate $\sqrt{-h} h^{ij} \delta \hat K_{ij}$ over $\partial {\cal M}$, any total divergence will contribute only a boundary term.    It is useful to use such integrations by parts to move all of the derivatives to act on $\epsilon_{ij}$.  Thus, we write
\begin{eqnarray}
\label{almostthere}
h^{ij} \delta \hat K_{ij} &=&  \frac{\rho}{2(d-2)} \delta {\cal R}_{ij} h^{ij} - \frac{1}{2} \hat K_{ij} \delta h^{ij}
+ \frac{\alpha }{2(d-2)\rho^{(d-5)}} \left[ D^2 \epsilon - D_i D_j \epsilon^{ij}
\right] \nonumber \\ &+& {\rm total \  divergence \ terms} \  + {\cal O}(\rho^{-(3d-8)}),
\end{eqnarray}
where the explicit terms on the final line of (\ref{xyz}) have summed to zero after applying (\ref{dhsub}).

At this stage, we need to express $\epsilon_{ij}$ in terms of
$h^1_{ij}$.  We may do so by treating $\frac{
h^1_{ij}}{\rho^{(d-5)}}$ as a perturbation of $\rho^2 h^0_{ij}$
and once again using (\ref{toSolve}) to solve perturbatively for
$\hat K_{ij}$ to the desired order.  The result simplifies greatly
when one uses the equations of motion.   It is at this stage that
we impose $d=4$, as the equations of motion for this case were
expanded in a convenient form in \cite{BS}.  In particular, we
make use of equations (3.27) from \cite{BS}, which allows us to
express the change $\Delta {\cal R}_{ij}$ in ${\cal R}_{ij}$
associated with changing the induced metric from $\rho^2 h^0_{ij}$
to $\rho^2 h^0_{ij} + \rho h^1_{ij}$ as
\begin{equation}
\label{DR}
 \Delta {\cal R}_{ij} = \frac{1}{2\rho} \left( 3
h^1_{ij} - h^1 h_{ij} + 2D_i D_j \sigma - 6 \frac{ \sigma
h_{ij}}{\rho^2} \right).
\end{equation}
Using this expression and (\ref{toSolve}), one may readily compute $\beta_{ij} = \hat K_{ij} - \rho h^0_{ij}$. Upon using
\begin{equation}
 (D^2 \sigma + 3 \sigma/ \rho^2) = 0,
\end{equation}
which is the equation of motion for $\sigma$ (equation (3.29) of \cite{BS}), 
we find
\begin{equation}
\label{solveE} \epsilon_{ij} = \beta_{ij} - h^1_{ij} = D_iD_j
\sigma - \frac{1}{2} h^1_{ij}.
\end{equation}

We may now compute:
\begin{equation}
D^2 \epsilon - D_i D_j \epsilon = D^4 \sigma - \frac{1}{2} D^2 h^1 - D_i D_j D^i D^j \sigma + \frac{1}{2} D^i D^j h^1_{ij}.
\end{equation}

Finally, we use the equation of motion (3.25) from \cite{BS},
\begin{equation}
-D^2 h^1 + D^i D^j h^1_{ij} = - \frac{12 \sigma}{\rho^4} + \ {\rm higher \ order \ terms},
\end{equation}
as well as their relation (A.2),
\begin{equation}
[D_i,D_j] \omega_k = h^0_{ki}\omega_{j} - h^0_{kj}\omega_{i}   + \ {\rm higher \ order \ terms},
\end{equation}
which follows from the commutator of covariant derivatives on the hyperboloid, to find
\begin{equation}
D^2 \epsilon - D_i D_j \epsilon = -\frac{2}{\rho^2} (D^2 \sigma + 3 \sigma/ \rho^2) = 0,
\end{equation}
Substituting this result into (\ref{almostthere}) yields (\ref{answer}), as desired.

\section{Sub-leading terms in the variation $\delta S_{\sqrt{\cal R}}$}
\label{C} 

This appendix outlines the calculation showing that $S_{renorm}$
as defined by the counter term $S_{\sqrt{\cal R}}$ is {\it not}
stationary for $d=4$ when one chooses both the temporal and
spatial cut-offs to be hyperbolic.  This result contrasts with the
result derived in appendix \ref{ErrorTerm}, showing that the
action defined by $S_{new \ CT}$ is indeed stationary.

We proceed by calculating (\ref{varSR}) to the next order in
$\rho$.  We begin by noting that
\begin{equation}
\delta \sqrt{\cal R} = \frac{1}{2} {\cal
R}^{-1/2}  \ \delta {\cal R}.
\end{equation}

Now, as in appendix \ref{ErrorTerm}, it is useful to compare
quantities such as ${\cal R}$ computed from the metric $h_{ij} =
\rho^2 h^0_{ij}$ with those computed from $h_{ij} = \rho^2
h^0_{ij} + \rho^{5-d} h^1_{ij}$.  Let us denote the corresponding
change in any such quantity by $\Delta$; i.e., $\Delta {\cal R}$
is the change in ${\cal R}$.  Then we have
\begin{equation}
\label{terms}  \delta \sqrt{\cal R} =
\frac{1}{2}\left(
\frac{\rho}{\sqrt{(d-1)(d-2)}} \left[h^{ij} \delta {\cal R}_{ij} +
{\cal R}_{ij} \delta h^{ij}\right] + \Delta {\cal R}^{-1/2}\delta
{\cal R} \right) .
\end{equation}

Let us now examine each term in turn.  Using (\ref{dR}),
the first term (involving $h^{ij} \delta {\cal R}_{ij}$) can be
written as a total derivative. Thus, it does not contribute.

To evaluate the second term to the desired order, we use
\begin{eqnarray}
{\cal R}_{ij} &=& (d-2) h^0_{ij} + \Delta {\cal R}_{ij} \cr &=&
\frac{(d-2)}{\rho} \hat K_{ij} - \frac{(d-2)}{\rho}D_iD_j \sigma -
\frac{(d-2)}{2\rho} h^1_{ij} + \Delta {\cal R}_{ij} ,
\end{eqnarray}
where in the last step we have used (\ref{solveE}). Specializing to 
$d=4$ and using (\ref{DR}) we have
\begin{equation}
{\cal R}_{ij} \delta h^{ij} \approx \frac{2}{\rho} \hat
K_{ij}\delta h^{ij} +  \frac{2\alpha}{\rho^{2}} D^2 \sigma +
\frac{\alpha }{\rho^{2}} h^1 + \frac{12 \alpha
\sigma}{\rho^{4}}  ,
\end{equation}
where the symbol $\approx$ indicates that we have dropped terms
which are high enough order in $\rho^{-1}$ that they will not
contribute to our final expression.  In addition, we have used
(\ref{dhsub}) and (\ref{DR}) in the subleading terms.

Finally, we compute the third term in (\ref{terms}).  From
(\ref{dR}) and (\ref{dhsub}) we find
\begin{equation}
\delta {\cal R} \approx - \frac{2}{\rho} (D^2 \alpha + 3 \alpha/
\rho^2). \end{equation} Furthermore, (\ref{DR}) yields
\begin{equation}
\Delta {\cal R} = - \frac{12\sigma}{\rho^3} - \frac{2}{\rho}
h^1.
\end{equation}
Thus, we have
\begin{equation}
\frac{1}{2} \Delta {\cal R}^{-1/2}\delta
{\cal R} \approx - \frac{\rho}{12} \alpha (D^2 h^1 + 3
h^1/ \rho^2) + \ {\rm total \ derivative \ terms},
\end{equation}
where we have also used the equation of motion (\ref{sigeq}).

Putting these results together we have
\begin{equation}
 \delta \sqrt{\cal R} = \frac{1}{\sqrt{6}}\left( \hat
K_{ij}\delta h^{ij} - \frac{\alpha\rho}{6} D^2 h^1
-+3\frac{\alpha\sigma}{\rho^3} \right)+ {\rm total \
derivative \ terms}
\end{equation}

Comparing with (\ref{answer}), we see that when the counter-term
$S_{\sqrt{\cal R}}$ is used, the variation $\delta S_{renorm}$
does not generically vanish on solutions.  However, it does vanish
on the covariant phase space defined by \cite{ABR}, where $h^1 =
-6 \frac{\sigma}{\rho^2}$.

\subsection*{Acknowledgments}

D.M. would like to thank Dumitru Astefanesei for conversations
which inspired this work. He would also like to thank Chris
Beetle, Jim Hartle, Marc Henneaux, Gary Horowitz, Rafael Sorkin and especially
Abhay Ashtekar for discussions of asymptotic flatness and
conserved quantities, as well as Vijay Balasubramanian, Jan de Boer, Per Kraus, and Kostas Skenderis
for other useful discussions and Julian le Witt and Simon Ross for catching several typos in an earlier draft. D.M. was supported in part by NSF
grant PHY0354978, by funds from the University of California, and
by funds from the Perimeter Institute of Theoretical Physics. D.M.
also wishes to thank the Perimeter Institute for their hospitality
during the initial stages of this work. R.M. was supported by the
Natural Sciences and Engineering Research Council of Canada.

\end{document}